\documentclass[prd,showpacs,amsmath,amssymb,nofootinbib]{revtex4}
\usepackage[dvips]{graphicx}
\usepackage{enumerate}
\newcommand \beq{\begin{eqnarray}}
\newcommand \eeq{\end{eqnarray}}

\def\simge{\mathrel{%
       \rlap{\raise 0.511ex \hbox{$>$}}{\lower 0.511ex \hbox{$\sim$}}}}
\def\simle{\mathrel{
       \rlap{\raise 0.511ex \hbox{$<$}}{\lower 0.511ex \hbox{$\sim$}}}}

\begin{document}

\title{Phase structure, collective modes, and the axial anomaly in dense QCD}
\author{Naoki Yamamoto,$^{1}$ Motoi Tachibana,$^{2}$ Tetsuo Hatsuda,$^{1}$
and Gordon Baym$^{3}$}
\affiliation{
$^{1}$Department of Physics, University of Tokyo, Japan\\
$^{2}$Department of Physics, Saga University, Saga 840-8502, Japan\\
$^{3}$Department of Physics, University of Illinois, 1110 W. Green St.,
Urbana, Illinois 61801}

\begin{abstract}

    Using a general Ginzburg-Landau effective Lagrangian, we study the
topological structure and low-lying collective modes of dense QCD having both
chiral and diquark condensates, for two and three massless flavors.  As we
found earlier, the QCD axial anomaly acts as an external field applied to the
chiral condensate in a color superconductor and, as a new critical point
emerges, leads to a crossover between the broken chiral symmetry and color
superconducting phases.  At intermediate densities where both chiral and
diquark condensates are present, we derive a generalized
Gell-Mann-Oakes-Renner relation between the masses of pseudoscalar bosons and
the magnitude of the chiral and diquark-condensates.  We show explicitly the
continuity of the ordinary pion at low densities to a generalized pion at high
densities.

\end{abstract}
\maketitle

\section{Introduction}

    Quantum chromodynamics at finite temperature, $T$, and chemical potential,
$\mu$, has a rich phase structure:  at low $T$ and $\mu$, the Nambu-Goldstone
(NG) phase with nearly massless pions is realized by the dynamical breaking of
chiral symmetry through condensation of quark$-$anti-quark pairs
\cite{NJL,HK94}.  On the other hand, at low $T$ and high $\mu$, a Fermi liquid
of deconfined quarks is expected to appear as a consequence of asymptotic
freedom \cite{QM}.  Furthermore, in cold quark matter, condensation of
quark$-$quark pairs \cite{CSC,RWA} leads to a color superconducting phase
(CSC) and dynamical breaking of color gauge symmetry.  At high $T$ for any
$\mu$, the condensates melt away and a quark-gluon plasma symmetric under
$SU(3)_C$ is realized \cite{YHM}.  The experimental exploration of quark-gluon
plasmas is being actively carried out in ultrarelativistic heavy ion
collisions at RHIC, and will be continued in the future at the LHC.  The
transition from the NG phase to the CSC is also relevant to heavy ion
collisions to be realized in the future at moderate energies at GSI, and is of
interest in the interiors of neutron stars and possible quark stars.

    We focus our attention here on the QCD phase structure at intermediate
densities and study the important interplay between two competing phenomena:
quark$-$anti-quark pairing, characterized by a chiral condensate $\Phi \sim
\langle \bar{q}q \rangle$, and quark$-$quark pairing, characterized by a
diquark condensate $d \sim \langle qq \rangle$.  This interplay is interesting
not only in its own right, but also in relation to similar phenomena in other
systems, e.g., the interplay between magnetically ordered phases and metallic
superconductivity \cite{SU91}.  Many works on the phase structure in the
intermediate density region have been carried out using effective models such
as that of Nambu and Jona-Lasinio (NJL) \cite{NJL-model} and the random matrix
model \cite{RMT}.  In Ref.~\cite{HTYB} we introduced a model-independent
approach to construct a general Ginzburg-Landau (GL) effective Lagrangian in
terms of the two order parameter fields, $\Phi$ and $d$, and showed, in
particular, that the $\Phi$-$d$ coupling induced by the axial anomaly leads to
a new critical point in the QCD phase diagram, implying a crossover between
the NG and CSC phases.  The crossover is relevant to the question of
``hadron-quark continuity" \cite{SW99}.

    The purposes of this paper are twofold:  Firstly, we give a detailed
account of the role of the axial anomaly in creating the topological structure
of the QCD phase diagram discussed in Ref.~\cite{HTYB}.  Secondly, we study
the properties of the collective modes, the pseudoscalar nonet ($\pi$, $K$,
$\eta$, $\eta'$) and the $H$ (the NG boson associated with the breaking of
$U(1)$ baryon symmetry), in the intermediate density regime where both chiral
and diquark condensates are present.

    This paper is organized as follows.  In Sec.~\ref{sec:phase}, we construct
a general GL free-energy in terms of the chiral and diquark fields that is
invariant under global chiral and local color symmetry.  We then use the GL
free energy to classify possible phases for two characteristic cases:
three-flavor quark matter with equal numbers of massless up, down and strange
quarks, and two-flavor quark matter with equal numbers of massless up and down
quarks; and then conjecture the phase diagram for realistic quark masses
(light up and down quarks and a medium-heavy strange quark) from the results
of the two cases.  In Sec.~\ref{sec:spectra}, after showing the essential
features of the mixing between independent massless modes by a toy model, we
build an effective Lagrangian for the ``pions" ($\pi$, $K$, $\eta$), and the
$H$ boson, and derive a generalized Gell-Mann-Oakes-Renner (GOR) relation in
dense QCD between the masses of the pseudo-scalar bosons and the magnitude of
the chiral and diquark-condensates.  We discuss the continuity of the NG phase
and CSC phase in the excited states.  Section~\ref{sec:summary} is devoted to
summary and conclusion.  In appendices, we give the detailed derivation of
the phase diagrams, magnitudes and signs of the GL free-energy parameters,
and the mass spectra of $\eta'$ mesons.

\section{Phase structure}
\label{sec:phase}

    In this section, we construct, on the basis of QCD symmetry, the most
general GL free-energy for the chiral field, the diquark fields and their
mutual couplings in three spatial dimensions.  We then classify the possible
phase structures of massless QCD with two and three flavors, and
discuss the QCD phase structure in the real world.

\subsection{Ginzburg-Landau free energy}

    We write the general GL free energy, measured with respect to that
of the normal phase, $\Phi=d_L=d_R=0$, as
\beq \label{Omega_full}
 \Omega(\Phi,d_L,d_R)
  = \Omega_{\chi}(\Phi)+\Omega_d(d_L,d_R) + \Omega_{\chi d}(\Phi,d_L,d_R),
\eeq
where $\Phi$, the chiral condensate, and $d_{_{L,R}}$, the diquark
condensates, are the basic order parameter fields; $\Omega_{\chi d}$ denotes
the coupling between these order parameters.  The form of
$\Omega(\Phi,d_L,d_R)$ is severely constrained by the QCD symmetry$^1$
\footnotetext[1]{More precisely, to avoid double counting the $Z(3)$ center in
$SU(3)$ and the discrete subgroup of $U(1)$, the chiral $SU(3)_{L,R}$'s
in Eq.~(\ref{eq:G-symmetry}) are the quotient groups, $SU(3)_{L,R}/Z(3)$.}
\beq
\label{eq:G-symmetry}
  {\cal G} \equiv SU(3)_L \times SU(3)_R \times U(1)_B \times U(1)_A \times
    SU(3)_C.
\eeq

    The axial anomaly breaks the $U(1)_A$ symmetry down to $Z(2N_f)_A=Z(6)_A$
where $N_f$ is the number of massless flavors.  Under ${\cal G}$, the left and
right handed quarks transform as
\beq
 q_L \rightarrow {\rm e}^{-i\alpha_A}\ {\rm e}^{-i\alpha_B} \ V_{L}\
  V_{C}\  q_L,
  \  \  \ \ q_R \rightarrow {\rm e}^{i\alpha_A} \ {\rm e}^{-i\alpha_B}\
  V_{R}\  V_{C}\  q_R,
\eeq
where $V_{L,R,C}$ are $SU(3)_{L,R,C}$ rotations with continuous
parameters and the phases $\alpha_{A,B}$ are associated with $U(1)_{A,B}$
rotations.

    We introduce the chiral fields $\Phi_{ij}$ (with $i$ and $j$ flavor
indices) of mass dimension one and singlets under $SU(3)_C$ and $U(1)_B$, as
\beq
\label{eq:phi-def}
   \Phi_{ij} =  G_\chi \langle\bar{q}_{Ra}^j q_{La}^i \rangle.
\eeq
The precise proportionality constant, $G_\chi$, enters only in numerical
evaluation of the GL free energy parameters; see Appendix \ref{sec:sign-gamma}.
The chiral fields transform under ${\cal G}$ as
\beq \label{trans:phi}
 \Phi \rightarrow {\rm e}^{-2i \alpha_A} \ V_{L} \ \Phi \ V_{R}^{\dagger} .
\eeq
In addition,
\beq \label{trans:phi2}
 \Phi \Phi^{\dagger} \rightarrow V_{L} (\Phi \Phi^{\dagger})V_{L}^{\dagger}, \ \
 \Phi^{\dagger}\Phi \rightarrow V_{R} (\Phi^{\dagger} \Phi)V_{R}^{\dagger}, \ \
  {\rm det}\ \Phi \rightarrow {\rm e}^{-6i \alpha_A} {\rm det}\ \Phi .
\eeq
Note the invariances of $\Phi$ under $Z(2)_A \subset U(1)_A$, and ${\rm
det}\ \Phi$ under $Z(6)_A \subset U(1)_A$.

    The most general form of the free energy of the chiral field, up to ${\cal
O}(\Phi^4)$, that is consistent with these transformations reads \cite{PW84},
\beq
   \label{eq:GL-chi}
   \Omega_{\chi} = \frac{a_0}{2} {\rm Tr}  \  \Phi^{\dagger} \Phi
   + \frac{b_1}{4!} \left( {\rm Tr} \ \Phi^{\dagger} \Phi    \right)^2
   + \frac{b_2}{4!} {\rm Tr} \left( \Phi^{\dagger} \Phi    \right)^2
     - \frac{c_0}{2} \left( {\rm det} \Phi + {\rm det} \Phi^{\dagger} \right),
\eeq
where ``Tr" and ``det" are taken over the flavor indices, $i$ and $j$.
The first three terms on the right side of Eq.~(\ref{eq:GL-chi}) are invariant
under the full symmetry group ${\cal G}$, while the fourth term, caused by the
axial anomaly, breaks the $U(1)_A$ symmetry down to $Z(6)_A$.  The free energy
$\Omega_{\chi}$ is bounded below for large values of the condensates for
$b_1+b_2/3>0$ and $b_2>0$.  If these conditions are not met, we need to
introduce terms ${\cal O}(\Phi^6)$ to stabilize the free energy, a situation
we will encounter later.  We assume $c_0$ to be positive so that the
condensate at low temperature is positive, while simultaneously the $\eta'$
mass obeys $m_{\eta'}^2 > 0$.  We assume that $a_0$ changes sign at a certain
temperature to drive the chiral phase transition.

   We define the  Lorentz-scalar diquark  order parameters,
\beq\label{phi-dl}
   G_d \langle q_{Lb}^j \ C\  q_{Lc}^k \rangle = \epsilon_{abc}
   \epsilon_{ijk} [d_{_L}^{\dagger}]_{ai} , \ \ \
   G_d \langle q_{Rb}^j \ C\  q_{Rc}^k \rangle = \epsilon_{abc}
     \epsilon_{ijk} [d_{_R}^{\dagger}]_{ai} ,
\eeq
where $i,j,k$ are the flavor and $a,b,c$ the color indices; $C=i \gamma^2
\gamma^0$ is the charge conjugation matrix with the properties, $CC^{\dagger}
= - C^2 = 1$ and $\gamma^0 C^{\dagger} \gamma^0 = C$.  The 3$\times$3 matrix
$[ d_{_{L,R}} ]_{ia}$ belongs to the {\em fundamental} representation of
$SU(3)_C \times SU(3)_{L,R}$.  The positive proportionality constant $G_d$, is
introduced to make the mass dimension of $d_{L,R}$ one; it enters only
in numerical evaluation of the GL free energy parameters (see Appendix
\ref{sec:sign-lambda}).  The transformation properties of the $d$'s under
${\cal G}$ are
\beq
\label{eq:dL-dR}
  d_L \to e^{2i \alpha_A}\  e^{2i \alpha_B}\  V_L \  d_L \ V_C^{\rm T},
  \ \ \ \ \ \
  d_R \to e^{-2i \alpha_A} \ e^{2i \alpha_B} \ V_R \ d_R \ V_C^{\rm T}.
\eeq
Note that $d_{L(R)}$ is invariant under $SU(3)_{R(L)}$, and $Z(2)_{B-A
(B+A)}$.  Also the color-singlet combinations of the diquark fields transform
as
\begin{align}
\label{trans:dLL}
  d_L d_L^{\dagger} & \rightarrow V_L \ (d_L d_L^{\dagger})\ V_L^{\dagger},&
  d_R d_R^{\dagger} &\rightarrow  V_R \ (d_R d_R^{\dagger})\ V_R^{\dagger},&
    \\
\label{trans:dLR}
   d_L d_R^{\dagger} & \rightarrow e^{4i \alpha_A}\ V_L \ (d_L
    d_R^{\dagger})\  V_R^{\dagger},&
  d_R d_L^{\dagger} &\rightarrow e^{-4i \alpha_A}\ V_R \ (d_R d_L^{\dagger})\
    V_L^{\dagger},&
\\
\label{trans:d-det}
    {\rm det} \ d_L & \rightarrow e^{6i \alpha_A}  e^{6i \alpha_B} {\rm
    det} \ d_L ,&
   {\rm det} \ d_R  &\rightarrow   e^{-6i \alpha_A} e^{6i \alpha_B} {\rm
    det} \ d_R .&
\end{align}
 Note that in addition to invariances under $SU(3)_C$, $U(1)_A$ and $U(1)_B$,
the products $d_L d_L^{\dagger}$ and $d_R d_R^{\dagger}$ are invariant under
$SU(3)_R$ and $SU(3)_L$ respectively.  Furthermore, in addition to invariances
under $SU(3)_C$, the products $d_L d_R^{\dagger}$ and $d_R d_L^{\dagger}$ are
invariant under $U(1)_B$ and $Z(4)_A \subset U(1)_A$, while ${\rm det} \ d_{L}$
and ${\rm det} \ d_{R}$ are invariant under $SU(3)_C$, $Z(6)_B \subset U(1)_B$,
and $Z(6)_A \subset U(1)_A$.

    The most general form of the GL free energy of the $d$-fields invariant
under $\cal G$ up to ${\cal O}(d^4)$ is \cite{IB,GR,IMTH}:
\beq
   \label{eq:GL-d}
   \Omega_{d} &=&
     \alpha_0 \ {\rm Tr} [d_L^{\ } d_L^{\dagger} +d_R^{\ }  d_R^{\dagger} ]
    + \beta_{1} \left(  [ {\rm Tr}(d_L^{\ } d_L^{\dagger})]^2
                        + [ {\rm Tr}(d_R^{\ } d_R^{\dagger})]^2    \right)
   \nonumber \\
   &+& \beta_{2}  \left(  {\rm Tr} [(d_L^{\ } d_L^{\dagger})^2]
                        + {\rm Tr} [(d_R^{\ } d_R^{\dagger})^2]    \right)
 + \beta_3 \ {\rm Tr} [(d_R^{\ } d_L^{\dagger})(d_L^{\ }d_R^{\dagger})]
 +  \beta_4 \ {\rm Tr} (d_L^{\ }d_L^{\dagger}){\rm Tr}(d_R^{\ }d_R^{\dagger}).
\eeq
The transition from the normal state to color superconductivity is driven by $\alpha_0$
changing sign.  Unlike for $\Phi$, terms such as ${\rm det}\ d_{L,R}$ are not
allowed, since $d_{L,R}$ carries baryon number.

    Finally, the interaction free-energy of the chiral and diquark fields to
fourth order reads \cite{HTYB}
\beq
\label{eq:GL-coup}
   \Omega_{\chi d}
   &= & \gamma_{1} \ {\rm Tr} [  (d_R^{\ } d_L^{\dagger})\Phi
                    + (d_L^{\ } d_R^{\dagger})\Phi^{\dagger}]
  + \lambda_{1} \ {\rm Tr} [(d_L^{\ } d_L^{\dagger})\Phi \Phi^{\dagger}
                  +(d_R^{\ } d_R^{\dagger})\Phi^{\dagger}\Phi ] \nonumber \\
  &&+\lambda_{2} \ {\rm Tr} [d_L^{\ } d_L^{\dagger} + d_R^{\ } d_R^{\dagger}]
                      \cdot {\rm Tr} [\Phi^{\dagger}\Phi ]
  + \lambda_{3} \left( {\rm det} \Phi \cdot
                    {\rm Tr}[(d_L^{\ } d_R^{\dagger}) \Phi^{-1}] + h.c
\right) .
\eeq
The triple boson coupling $\sim \gamma_1$, which breaks the $U(1)_A$
symmetry down to $Z(6)_A$, originates from the axial anomaly.  The remaining
terms are fully invariant under $\cal G$.  The $\lambda_3$ term is equivalent
to the polynomial structure, $\epsilon_{ijk} \epsilon_{i'j'k'} \Phi_{ii'}
\Phi_{jj'} (d_L d_R^{\dagger})_{kk'}$ \cite{SS00}.

    Equations (\ref{eq:GL-chi}), (\ref{eq:GL-d}) and (\ref{eq:GL-coup})
constitute the most general form of the GL free energy under the conditions
that the phase transition is not strongly first order (i.e., the values of
$\Phi, d_{L,R}$ are sufficiently smaller than those at zero temperature) and
that the condensed phases are spatially homogeneous.

\subsection{Three massless flavors}
\label{ssec:massless-3}

    We first study three massless flavors.  To proceed analytically, we
restrict ourselves to maximally symmetric condensates, namely, a flavor
symmetric chiral condensate in which quark$-$anti-quark pairing takes place
only within the same flavor and a color-flavor-locked (CFL) diquark condensate
in which the quark-quark pairing takes place only in the different flavors:
\beq
\label{DIA ansatz}
 \Phi &=&{\rm diag}(\sigma, \sigma,\sigma), \\
 \label{CFL ansatz} d_L &=&-d_R={\rm diag}(d,d,d),
\eeq
where $\sigma$ and $d$ are assumed to be real.  We have chosen the
relative sign between $d_L$ and $d_R$ in Eq.~(\ref{CFL ansatz}) corresponding
to the ground state having positive parity, as is indeed favored by the axial
anomaly together with finite quark masses \cite{TS}.  The CFL condensate in
Eq.~(\ref{CFL ansatz}) is indeed realized at asymptotically high density, as
shown in \cite{IB} via the GL approach, and is the simplest ansatz at
intermediate density.  Thus the reduced GL free energy becomes
\beq
    \label{eq:nf3-model}
    \Omega_{3F}(\sigma,d)
    &=& \left( \frac{a}{2} \sigma^2 - \frac{c}{3} \sigma^3 +
    \frac{b}{4}\sigma^4
    + \frac{f}{6}\sigma^6 \right)
    + \left( \frac{\alpha}{2} d^2 + \frac{\beta}{4} d^4 \right)
     - {\gamma} d^2 \sigma +  {\lambda} d^2 \sigma^2.
\eeq
Some comments are in order on the coefficients of the terms in
$\Omega_{3F}(\sigma,d)$:
\begin{enumerate}

    \item[(i)] Changes in the magnitude of $a$ and $\alpha$ drive the phase
transitions.  In the following, we use these variables as the key parameters
in classifying the phase structure.

    \item[(ii)] As suggested in effective theories and in lattice QCD
simulations \cite{AY89}, $b$ may change sign as a function of $T$ and $\mu$
(see Eq.~(\ref{eq:b-integral}) in Appendix \ref{sec:sign-lambda}).
To incorporate such a possibility,
we have introduced a ${\cal O}(\sigma^6)$ term with a
coefficient $f >0$ to stabilize the system.  As we will see later, the
qualitative phase structure for three massless flavors is independent of the
sign of $b$.

    \item[(iii)] We assume that $\beta$ is always positive, as expected from
effective theories and weak coupling QCD (see Eq.(\ref{eq:beta-integral}) in
Appeidix \ref{sec:sign-lambda}).
This sign implies that the transition from the normal phase to CSC
is second order at the tree-level if there is no coupling between $d$ and $\Phi$.
Fluctuations of the diquark or gluon fields can make the transition first
order \cite{MIHB}, an effect we will not treat in this paper.

    \item[(iv)] Because both the $d^2 \sigma$ and $\sigma^3$ terms in
Eq.~(\ref{eq:nf3-model}) originate from the axial anomaly, their coefficients,
$\gamma$ and $c$, are related microscopically.  Indeed, it can be shown from
the instanton-induced six-fermion interaction, $\sim\det_{i,j}(\bar{q}_R^j
q_L^i)$, that $\gamma$ has the same sign and the same order of magnitude as
$c$ (Appendix \ref{sec:sign-gamma}).  Since $c$ is positive, as noted below
Eq.~(\ref{eq:GL-chi}), $\gamma$ is also positive.  By lowering the free
energy, the $\gamma$-term favors coexistence:  $\sigma \neq0$ and $d \neq 0$,

    \item[(v)] Microscopic calculations in weak coupling QCD and in the NJL
model (Appendix \ref{sec:sign-lambda}) show that the coefficient $\lambda$ is
always positive.  Physically a non-vanishing $\sigma$ plays the role of an
effective mass for the quark field, reducing the density of states at the
Fermi surface, and the pairing energy \cite{IMTH}, an effect represented by
$\lambda d^2 \sigma^2 >0$.  Furthermore, as shown in Appendix
\ref{sec:sign-lambda}, $\lambda/\beta \sim \ln(\Lambda/T_d)/(\mu/T_d)^2$,
which is rather small for reasonable values of $\mu$, $T_d$ (the critical
temperature of the color superconductivity without the $\sigma$-$d$ coupling),
and $\Lambda$ ($\sim \mu$ for weak coupling QCD, and in the NJL model, of
order the spatial momentum cutoff).  Therefore, we focus primarily on the case
$\gamma >0$ with $\lambda =0$ for the three-flavor case.  The effect of the
small positive $\lambda$ for three-flavor case is discussed in Appendix
\ref{sec:phase-diagram}.

\end{enumerate}

    In principle, the system can have four possible phases:
\beq
    {\rm Normal\ (NOR)\  phase}      &: &  \sigma=0, d=0, \nonumber \\
    {\rm CSC\  phase}                &: &  \sigma=0, d\neq 0,\nonumber \\
    {\rm NG\ phase}                  &: &  \sigma\neq 0, d= 0, \nonumber \\
    {\rm Coexistence\ (COE) \ phase} &: &  \sigma\neq 0, d \neq 0.
\label{eq:phase-def}
\eeq

    The symmetry breaking pattern of the individual phases are the following:
in the CSC phase, ${\cal G} \to SU(3)_{C+L+R} \times Z(2)$; in the NG phase,
${\cal G} \to SU(3)_C \times SU(3)_{L+R} \times U(1)_B$, and in the COE phase,
${\cal G} \to SU(3)_{C+L+R} \times Z(2)$.  Note that the CFL phase and COE
phase with the diquark condensate of Eq.~(\ref{CFL ansatz}) each break the
original $U(1)_B \times Z(6)_A$ symmetry with axial anomaly down to $Z(2)$
(the simultaneous reflection, $q_{L,R} \rightarrow - q_{L,R}$), and thus these
two phases are not distinguished by symmetry.

\subsubsection{Outline of the phase diagram}

    The phase diagram in the $a$-$\alpha$ plane can be determined uniquely by
comparing the global minima of the free energies, $\Omega^{\rm (NOR)}$,
$\Omega^{\rm (CSC)}$, $\Omega^{\rm (NG)}$, and $\Omega^{\rm (COE)}$.  In the
following, we outline the basic aspects of the phase diagram obtained from
comparison of the free energies, and give the calculations of the phase
boundaries and critical points in detail in Appendix \ref{sec:phase-diagram}.

\begin{figure}[t]
\begin{center}
\includegraphics[width=8.0cm]{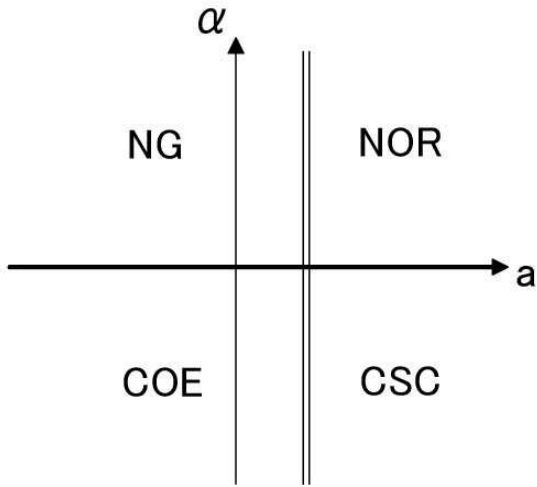}
\end{center}
\vspace{-1.0cm}
\caption{Phase structure in three and two-flavor systems without coupling
between the chiral and diquark condensates ($\gamma=\lambda=0$).  Phase
boundaries with a second order transition are denoted by a single line, and
with a first order transition by a double line.}

\label{fig:normal}
\end{figure}

    In the absence of the $\sigma$-$d$ coupling ($\gamma=\lambda=0$), the four
phases in (\ref{eq:phase-def}) are separated by $\alpha=0$ (a line of second
order transitions) and by $a=a_{\chi}$ (a line of first order phase
transitions), shown in Fig.\ref{fig:normal}, where the first and second order
phase boundaries are drawn as double and single lines, respectively.

\begin{figure}[t]
\begin{center}
\includegraphics[width=7.5cm]{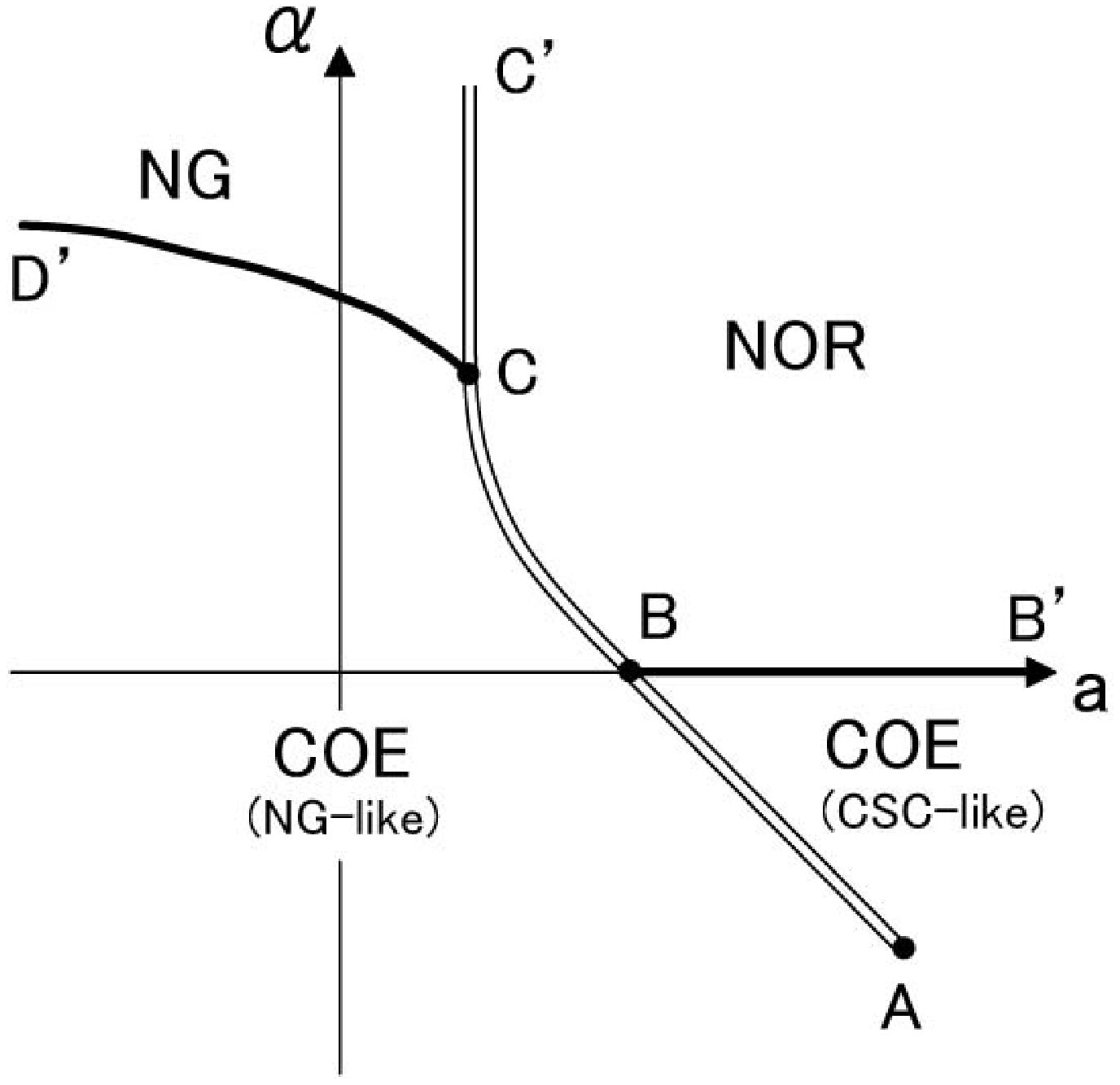}
\includegraphics[width=7.5cm]{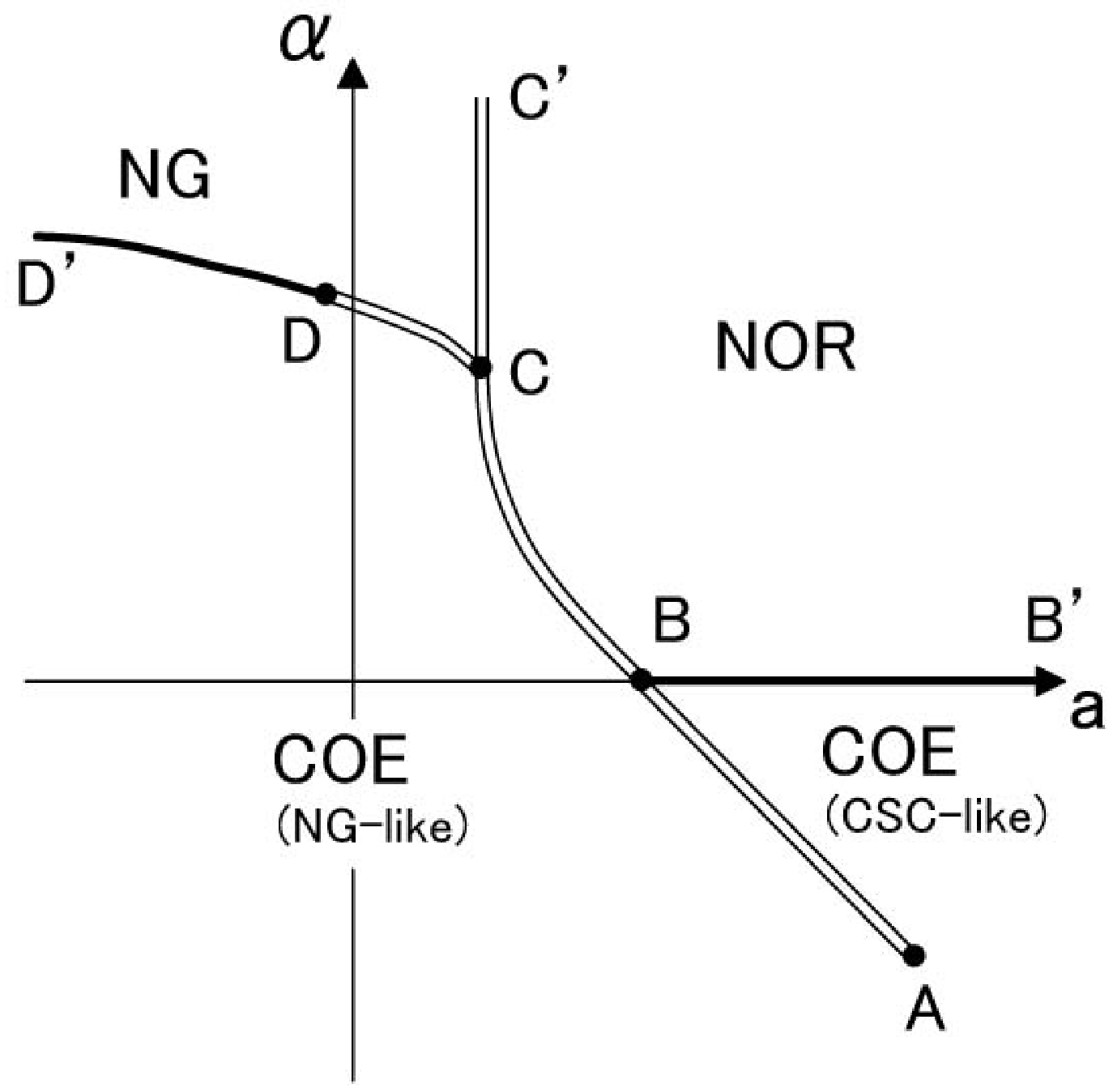}
\end{center}
\vspace{-0.5cm}
\caption{Phase structure in the three-flavor system with $\gamma >0$ and
$\lambda=0$.  Phase boundaries with a second order transition are denoted by a
single line, and with a first order transition by a double line.  On the left,
C is a critical end point ($0<\gamma < (c/3)\sqrt{\beta/b}$), and on the
right, a triple point ($\gamma > (c/3)\sqrt{\beta/b}$).
}
\label{fig:3F}
\end{figure}

    An {\it attractive} coupling, $\gamma > 0$ (with $\lambda=0$) leads to
several important modifications, summarized in (i)-(iv) below and shown in
Fig.\ref{fig:3F} (the left panel for small $\gamma$ and the right panel for
large $\gamma$).$^2$\footnotetext[2]{In describing the phase diagram we denote
a point where a first order line turns into a crossover as a {\em critical
point}; a point where a first order line turns into a second order line as a
{\em tricritical point}; a point where a second order line terminates on a
first order line as a {\em critical end point}; the intersection of three
first order lines as a {\em triple point}; and a point where two or four
second order lines meet as a {\em bicritical} or {\em tetracritical point}
respectively \cite{CL}.}

\begin{enumerate}

    \item[(i)] The area of the COE phase grows when both $\sigma$ and $d^2$
are non-vanishing and positive, since the $- \gamma d^2 \sigma$ term lowers
the free energy.

    \item[(ii)] The first order line between the COE and CSC phases,
originally the double vertical line in Fig.~\ref{fig:normal}, terminates at a
critical point, A, as shown in Fig.~\ref{fig:3F}.  This behavior is
anticipated, since $-\gamma d^2 $ acts as an external field for $\sigma$ and
washes out the first order phase transition for sufficiently large $\gamma$ or
$d$.  In Fig.~\ref{fig:3F} we denote the coexistence region contacting the NG
phase as the {\em NG-like COE phase}, and the region which was originally the
CSC phase as the {\em CSC-like COE phase}.

    \item[(iii)] The second order phase boundary originally located at
$\alpha=0$ splits in two, a line going to the right from the critical end
point B and a line going to the left from the point C. Since $\sigma$ changes
discontinuously across the first order phase boundary CB, the $d^2 \sigma$
term, which acts as a mass term for $d^2$, leads to different critical
temperatures for diquark condensation in the NG-like COE and CSC-like COE
phases.

    \item[(iv)] For large enough $\gamma$, a tricritical point D appears on
the boundary between the NG and COE phases.  Then the point C, otherwise a
critical end point, becomes a triple point.

\end{enumerate}

\subsection{Massless two-flavor case}
 \label{ssec:2-flavor}

    We turn now to the massless two-flavor system (with infinite strange quark
mass).  In this case, all chiral and diquark condensates with a strange quark
are suppressed. We write
\beq
  \Phi& =&{\rm diag}(\sigma,\sigma,0), \\
  d_L&=&-d_R={\rm diag}(0,0,d).
\eeq
The latter is the two flavor color superconductivity (2SC) state.  Due to
this color-flavor structure, the cubic terms in $\sigma$ and $d$ are
identically zero, and the model reduces to:
\beq
\label{eq:nf2-model}
\Omega_{2F}(\sigma,d) =
   \left( \frac{a}{2} \sigma^2 + \frac{b}{4} \sigma^4
  + \frac{f}{6} \sigma^6 \right)
  +\left(  \frac{\alpha}{2} {d}^2 + \frac{\beta}{4} {d}^4 \right)
  + {\lambda} {d}^2 \sigma^2.
\eeq

    We locate the phase boundaries and the order of the phase transitions by
comparing the free energies:  $\Omega^{\rm (CSC)}(0,d) = \frac{\alpha}{2}d^2 +
\frac{\beta}{4}d^4$, $\Omega^{\rm (NG)}(\sigma,0) = \frac{\alpha}{2} \sigma^2
+ \frac{b}{4}\sigma^4 + \frac{f}{6}\sigma^6$ and $\Omega^{\rm
(COE)}(\sigma,d)$, with respect to $\Omega^{\rm (NOR)}(0,0)$.

\subsubsection{Outline of the phase diagram}

\begin{figure}[t]
\begin{center}
\includegraphics[width=13.0cm]{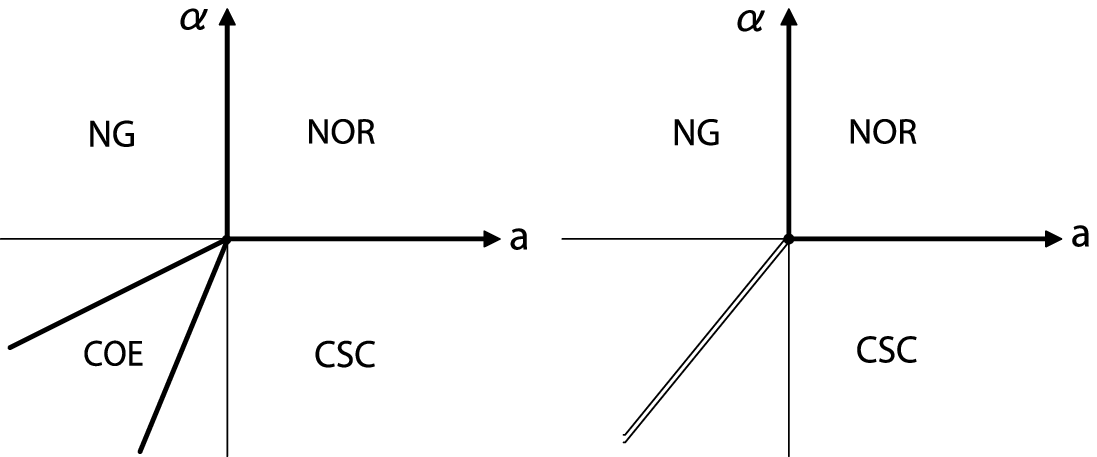}
\end{center}
\vspace{-0.5cm}
\caption{Phase strucuture in the two-flavor system for $b>0$ and $\lambda
> 0$.  Left:  Case with a tetracritical point ($\frac{1}{2}\sqrt{b\beta} >
\lambda > 0$).  The second order line between NG and COE is characterized by
$\alpha = 2a\lambda/b$, and that between CSC and COE by $\alpha =
a\beta/2\lambda$.  Right:  Case with a bicritical point ($\lambda >
\frac{1}{2}\sqrt{b\beta}$).  The first order line between NG and CSC is
characterized by $\alpha = a\sqrt{\beta/b}$.}
\label{fig:2Fp}
\end{figure}

    We first consider $b >0$ in Eq.~(\ref{eq:nf2-model}), with the $\sigma^6$
term safely neglected, to find the qualitative phase structure.  The system is
equivalent to an anisotropic antiferromagnet (e.g., GdAlO$_3$) \cite{CL}.  For
$\lambda=0$, the boundaries of the four phases in (\ref{eq:phase-def}) are
characterized by second order lines at $\alpha=0$ and $a=0$ with a
tetracritical point at $\alpha=a=0$.  In the presence of the {\it repulsive}
$d^2 \sigma^2$ term ($\lambda > 0$), the area of the coexistence phase
decreases, as shown in the left panel of Fig.~\ref{fig:2Fp}.  For $ \lambda >
\frac{1}{2}\sqrt{b\beta}$, the coexistence region disappears altogether, and a
first order interface between CSC and NG appears at $\alpha =
a\sqrt{\beta/b}$.  In addition, $a=\alpha=0$ becomes a bicritical point, as
shown in the right panel of Fig.~\ref{fig:2Fp}.

\begin{figure}[t]
\begin{center}
\includegraphics[width=7.0cm]{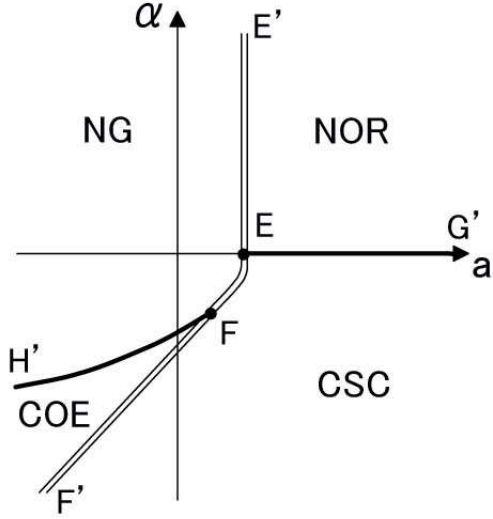}
\end{center}
\vspace{-0.5cm}
\caption{Phase strucuture of the two-flavor system with $b<0$ and $\lambda
> 0$.  The second order line between NG and COE is given by $\alpha=
\lambda(b-\sqrt{b^2-4fa})/f$.  The first order line between CSC and COE is
given by $\alpha=(\beta/2\lambda) [a- 3(b- 4\lambda^2/\beta)^2/(16f)]$, while
the first order line between NG and CSC is given by $\alpha^2 = \beta[6bfa +
(b^2-4fa)^{3/2}-b^3]/(6f^2)$.}
\label{fig:2Fn}
\end{figure}

    Next we consider $b <0$; here the $\sigma^6$ term plays an essential role.
For $\lambda=0$, the four phases in Eq.~(\ref{eq:phase-def}) are separated by
a second order line at $\alpha=0$ and a first order line at $a=3b^2/(16f)$ as
shown in Fig.~\ref{fig:normal}.  With a {\it repulsive} $d^2\sigma^2$ term
($\lambda > 0$) the coexistence phase shrinks and gradually fades away as
$\lambda \to \infty$.  Moreover, a new first order line between NG and CSC
appears and grows as $\lambda$ increases.  This situation is shown in
Fig.~\ref{fig:2Fn}.  Such a phase structure was previously pointed out by
Vanderheyden and Jackson \cite{RM-model} using the random matrix model, and by
Kitazawa et al.  \cite{NJL2-model} using the NJL model; our model-independent
analysis is consistent with these results.  However, the random matrix model
does not include the physics of the suppression of the density of states at
the Fermi surface leading to $\lambda > 0$, and thus its predicted phase
structure is not obviously related to that discussed here.

\subsection{Speculative phase structure for realistic quark masses}

\begin{figure}[h]
\begin{center}
\includegraphics[width=11.0cm]{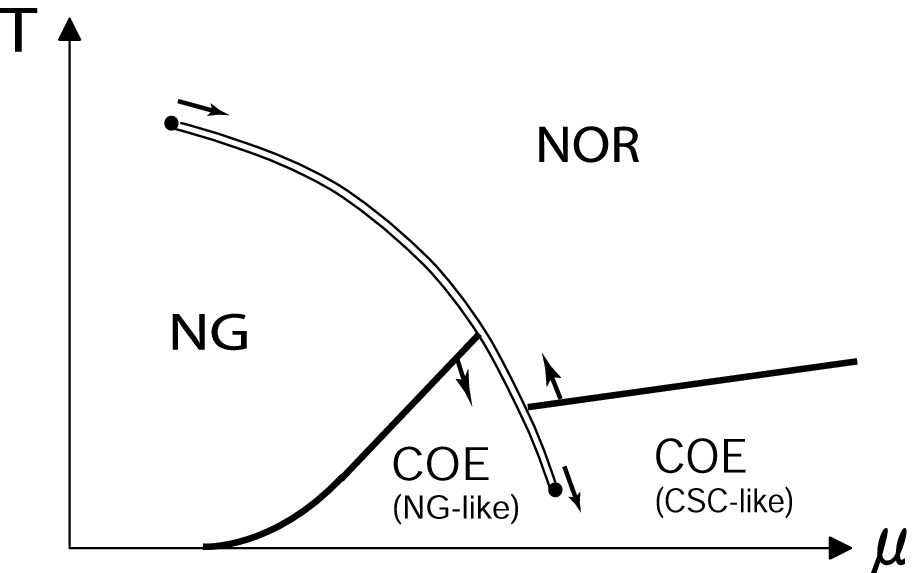}
\end{center}
\vspace{-0.5cm}
\caption{Schematic phase structure with two light (up and down) quarks and
a medium heavy (strange) quark.  The arrows show how the critical point and
the phase boundaries move as the strange-quark mass increases towards the
two-flavor limit.}
\label{fig:32F}
\end{figure}

    The mapping of the phase diagrams obtained in the ($a,\alpha$) plane to
the ($T,\mu$) plane is a dynamical question which we cannot address within the
phenomenological GL theory.  Nevertheless, on the basis of the results of
Sec.~\ref{sec:phase}, we can draw a {\em speculative} phase structure (one of
many possible structures) in Fig.~\ref{fig:32F} for two light (up and down)
quarks and a medium-heavy (strange) quark.

    In the figure, we draw two critical points:  the usual one near the
vertical axis (the high temperature critical point) originally found by
Asakawa and Yazaki \cite{AY89}, and a new one near the horizontal axis (the
low temperature critical point) driven by the axial anomaly.  We have assumed
that both critical points are in the region of positive $T$ and $\mu$, and
thus the chiral transition is a crossover in the direction of both high $T$
and high $\mu$.  With decreasing strange quark mass, $m_s$, the high-$T$
critical point approaches the vertical axis because the quark mass, which
tends to weaken the first order transiton, becomes less effective.  On the
other hand, as $m_s$ increases, the low-$T$ critical point approaches the
horizontal axis since the system approaches the two-flavor case where the
anomaly-induced critical point does not appear.  Whether this scenario is
realized or not needs to be checked in effective models incorporating the
axial anomaly, and eventually be checked by first principles QCD simulations.
Such calculations are beyond the scope of this paper.

    We should mention a similar two-critical-point phase structure found in
the two-flavor NJL model \cite{NJL2-model}.  Since the axial anomaly does not
produce a triple boson coupling in two flavors, as discussed in
Sec.~\ref{ssec:2-flavor}, the origin of the low-$T$ critical point in Ref.
\cite{NJL2-model} is not related to ours, but can be understood as follows:
Expanding the free energy of the two-flavor NJL model up to ${\cal O}(d^2
\sigma^4)$. we arrive at an ``effective" quartic term in the GL free energy,
\beq
 \frac{1}{4}  b \sigma^4 + \frac{1}{4} \tilde{b} d^2 \sigma^4 .
\eeq
It can be shown that $b$ changes sign from positive to negative, and
$\tilde{b}$ from negative to positive, as $\mu$ increases from zero.  (See
Eqs.~(\ref{eq:b-integral} and \ref{eq:bp-integral}) in Appendix \ref{sec:sign-lambda}.)
Thus for sufficiently large $\mu$ and $d$, the first order transition driven by $b <0 $
disappears due to $\tilde{b} d^2 >0$.  This implies that the critical point in
\cite{NJL2-model} in the two-flavor case originates from a large higher order
term in the GL expansion, and is thus beyond the reach of our strict GL
expansion.

\section{Excitation spectra of mesons}
\label{sec:spectra}

    We turn, in this section, to study the mass spectra of mesons including
$H$ boson in the intermediate density region for the degenerate three-flavor
case.  (The $\eta'$ is discussed in Appendix \ref{sec:spectra-eta'}.)  We
focus on the region near the phase boundaries where $\sigma$ and $d$ are
sufficiently small in the COE phase, and consider energy scales, $p$, smaller
than the pairing gap $d$, so that we can neglect excitations of the quarks.

    The symmetry breaking pattern for the chiral field at low density is
$SU(3)_L \times SU(3)_R \rightarrow SU(3)_V$.  Accordingly one has eight
Nambu-Goldstone (NG) bosons, identified with the pseudoscalar $\bar q q$-state
octet.  These pions are massless in the chiral limit, with masses generated by
finite quark masses.  (The $\eta'$ meson is much heavier than the pions
because of the $U(1)_A$ axial anomaly.)

    On the other hand, the pattern of symmetry breaking for the diquark fields
in the asymptotic high density limit is $SU(3)_C \times SU(3)_L \times SU(3)_R
\times U(1) _B \rightarrow SU(3)_{C+L+R}\times Z(2)_B$, leading altogether to
17 NG bosons.  Eight of these bosons are ``eaten" by the gluons through the
Anderson-Higgs mechanism; the remaining nine are the pseudoscalar octet
$\tilde{\pi}$ (the ``generalized pions," or $\bar q \bar q qq$-states as
discussed in \cite{RWA,SS00,F04,CG99,RWZ00}), together with the $H$
boson.$^3$\footnotetext[3] {In addition, one also has a massive
$\tilde{\eta}'$, since $U(1)_A$ remains broken by the axial anomaly in the
high density limit.  However, the violation of $U(1)_A$ symmetry becomes much
smaller as screening suppresses the instanton density by a high power of
$1/\mu$ \cite{TS}.  Correspondingly we have a light $\eta'$ meson at
asymptotically high density.}

    In the intermediate density region, two sets of light pseudoscalar mesons
are present, one arising from the chiral condensate ($\bar q q$-states) and
the other from the diquark condensates ($\bar q \bar q qq$-states); these two
sets are mixed through the interaction terms \cite{F04}.  Specifically, we
have two sets of pion octets ($\pi$ and $ \tilde \pi$) (as well as $\eta'$ and
$\tilde \eta'$) from the chiral and diquark fields, and in addition the $H$
boson from the diquark field in this region.  As we show below, mixing causes
one of the two pions to become heavy while the other becomes massless.  (A
similar discussion of the mixing between $\bar q q$- and $\bar q \bar q
qq$-states at $T=0$ and $\mu = 0$ is given in \cite{FJS06}.)  The $H$ boson
does not experience mixing, since it exists only in the diquark sector, but
remains exactly massless independent of the quark masses.

    In the following, we examine the mass spectra corresponding to this
physical picture more quantitatively using an effective Lagrangian approach.
We begin with a toy model with independent massless fields to show explicitly
the essential features of the role mixing plays.  Next we construct an
effective Lagrangian involving the pion octet appropriate in the intermediate
density region.  Using the nonlinear chiral Lagrangians studied for the NG
phase \cite{GL85} and for the CFL phase \cite {SS00,CG99,HLM00,MT00}, we take
into account $\pi$ and $ \tilde \pi$ and study the effects of the mixing terms 
determined in Sec.~\ref{sec:phase}, which were previously not considered.
The diagonalization of the mixing matrix for $\pi$ and $ \tilde \pi$ leads to a
generalized Gell-Mann-Oakes-Renner (GOR) relation in dense QCD, which relates 
the masses of pseudoscalar bosons to the magnitude of the chiral and diquark
condensates.  We also discuss the continuity between the NG and the CSC phases
associated with the excited states.

\subsection{A toy model}
\label{ssec:toy}

    To show the effect the mixing of independent fields in the presence of
mass terms, we first consider a simple toy model for two complex scalar
fields, $\alpha$ and $\tilde \alpha$:
\beq
 \label{toy model}
 {\cal L} &=& {\cal L}_{\rm kin}+{\cal L}_ {\rm mix}+{\cal L}_{\rm mass},
 \nonumber \\
 {\cal L}_{\rm kin} &=& \frac{f^2}{2}\left| {\partial _\mu  \alpha  }
 \right|^2  +
 \frac{\tilde f^2}{2}\left| {\partial _\mu
 \tilde \alpha } \right|^2,
 \nonumber \\
 {\cal L}_{\rm mix} &=& \frac{A}{2}\left( {\alpha \tilde \alpha ^ \dag   +
 {\rm h.c.} } \right),
 \nonumber \\
 {\cal L}_{\rm mass} &=& \frac{m}{2} \left(B \alpha  + C \tilde \alpha +{\rm
  h.c.} \right),
\eeq
where $A$, $B$, $C$, $f$, $\tilde f$ and $m$ are positive constants.

    We focus on the low mass phase excitations, described by the fields
$\varphi$ and $\tilde \varphi$ defined by
\beq \alpha=\exp\left({i\varphi/f}\right), \ \
  \tilde \alpha=\exp\left({i\tilde \varphi/\tilde f}\right).
\eeq
In the absence of the mixing term ${\cal L}_{\rm mix}$ and mass term
${\cal L}_{\rm mass}$, both $ \varphi$ and $\tilde \varphi$ are massless.  We
turn on ${\cal L}_{\rm mix}$ and ${\cal L}_{\rm mass}$, and expand the
Lagrangian in terms of $\varphi$ and $\tilde \varphi$.  After subtracting a
constant part, we obtain
\beq
 {\cal L}_{\rm mix}  +  {\cal L}_{\rm mass} & = &
 - \frac{1}{2}\left( {\begin{array}{*{20}c}
     {\varphi} & {\tilde \varphi }
 \\
 \end{array}} \right)M_{(\varphi)} \left ( {\begin{array}{*{20}c}
     {\varphi }  \\
     {\tilde \varphi }  \\
 \end{array}} \right)+{\cal O}(\varphi^4),
 \nonumber \\
  M_{(\varphi )} & = & \left( {\begin{array}{*{20}c}
    {{(A+mB)}/{f^2}} & { - {A}/{f \tilde f}}  \\
    { - {A}/{f \tilde f}} & {(A+mC)}/{{\tilde f}^2}  \\
 \end{array}} \right).
\eeq
The matrix $M_{(\varphi )}$ is diagonalized by a transformation
\beq
\label{toy:mixing}
\left( {\begin{array}{*{20}c}
    {\varphi _1 }  \\
    {\varphi_2 }  \\
\end{array}} \right) = \left( {\begin{array}{* {20}c}
    {\cos \Theta } & {\sin
\Theta }  \\
    { - \sin \Theta } & {\cos
\Theta }  \\
\end{array}} \right)\left( {\begin{array}{*{20} c}
    \varphi   \\
    {\tilde \varphi }  \\
\end{array}} \right),
\eeq
where $\Theta$ is the mixing angle; we thus we obtain the eigenmasses
$m_{\varphi_{1} }^2$ and $m_{\varphi_{2} }^2$ $(m_{\varphi_ {1} }^2 <
m_{\varphi_{2} }^2)$:  \\

\noindent (i) For $m=0$,
we have
\beq
  m_{\varphi_{1} }^2(m=0)  &=& 0 , \\
  m_{\varphi_{2} }^2(m=0)
   &=& A \left( \frac{1}{f^2}+\frac{1}{{\tilde f}^2} \right) .
\eeq
The presence of a massless excitation corresponds to simultaneous rotation
of $\varphi_1 $ and $\varphi_2$ in the same direction, while the massive mode
corresponds to rotation in opposite directions.
\\

\noindent (ii) For small but non-vanishing $m$, we obtain
\beq
 \label{toy:varphi1}
 m_{\varphi _{1} }^2 &\simeq&
  \frac{m}{ f^2  +  {\tilde f} ^2 } \left(B+C\right) ,
\\
\label{toy:varphi2}
  m_{\varphi _{2} }^2 &\simeq &
   m_{\varphi _{2} }^2(m=0) +
   \frac{m}{f^2 + {\tilde f}^2 }\left(
  B\frac{{\tilde f}^2 }{f^2 } + C \frac{f^2}{{\tilde f }^2} \right),
\eeq
In Sec.~\ref{sec:mass_spectra} we will discuss mass relations of this kind
arising via mixing in the realistic situation of QCD.  In particular,
Eq.~(\ref{toy:varphi1}) is a key relation corresponding to our later generalized
Gell-Mann-Oakes-Renner relation, Eq.~(\ref{mGOR-relation}).

\subsection{General effective Lagrangian}

    To derive the effective Lagrangian for the light excited states, the pions
($\pi$, $K$ and $\eta$), $\eta'$ and the $H$ boson in the intermediate density
region, we fix, as above, the magnitude of the chiral and diquark fields and
consider only fluctuations of their phases about their vacuum configurations.
We thus parametrize the fields as
\beq
 \label{parametrization}
 \Phi  = \sigma \Sigma  e^{ - 2i\theta },
 \ \ d_L  = dU_L e^{2i(\tilde \theta + \phi) },
 \ \ d_R  =  - dU_R e^{ - 2i(\tilde \theta  - \phi) },
\eeq
where $\Sigma$, $U_L$ and $U_R$ are $SU(3)$ matrices, and the angles
$\theta$ and $\tilde \theta$ are $U(1)_A$ phases, and $\phi$ the $U(1)_B$
phase.

    As mentioned at the top of Sec.~\ref{sec:spectra}, all eight gluons
acquire a mass of order $g f_{\tilde{\pi}} \sim {\cal O}(g \mu)$ by ``eating"
the eight colored fluctuations of $U_{L,R}$, where $g$ is the QCD coupling
constant and $f_{\tilde{\pi}}$ is the decay constant associated with
$U_{L,R}$, defined below.  On the low momentum scales we consider, $p < d \ll
gf_{\tilde \pi}$, gluons are not low-lying modes.  The remaining color-singlet
fluctuations correspond to $\tilde{\pi}$ and are parametrized by the field
\beq
 \label{new-field}
 \tilde \Delta = U_L U_R^ \dag ,
\eeq
which transforms under $SU(3)_L \times SU(3)_R$ as
\beq
 \tilde \Delta \rightarrow V_L \tilde \Delta V_R^\dag .
 \eeq

\subsubsection{Kinetic terms}

    The kinetic term invariant under ${\cal G}$ for the chiral fields
$\Sigma$ and $\theta$, to second order in derivatives is,
\beq
 \label{chi-kin-term}
 {\cal L}_{\chi} ^{\rm kin} = {f_\pi ^2} g_{\pi}^{\mu \nu }
 {\rm Tr}\left( \partial _\mu \Sigma \partial _\nu
 \Sigma ^\dagger  \right) + \frac{f_ {\eta'}^2}{2} {g_{\eta'}^{\mu \nu }
\partial _ \mu  \theta \partial _\nu  \theta },
\eeq
where the metric tensors
$g_{\pi}^{\mu \nu } = {\rm diag} (1,v_\pi^2,v_\pi^2,v_\pi^2)$
and
$g_{\eta'}^{\mu \nu } = {\rm diag} (1,v_{\eta '}^2,v_{\eta '}^2,v_{\eta '}^2)$
arise due to the absence
of Lorentz invariance in the medium; Here $v_{\pi}$ and $v_{\eta '}$ are the
speeds of the pions and $\eta'$, respectively.

    The standard pion fields $\pi^j$ are defined by
\beq
  \Sigma  = \exp \left(i\lambda^j \pi^j/{f_\pi} \right),
\eeq
where the $\lambda^j$ $(j=1,\ldots,8)$ are Gell-Mann matrices normalized
so that Tr$\lambda^i \lambda^j=\frac{1}{2}\delta ^{ij}$.  The first term in
Eq.~(\ref{chi-kin-term}) is the standard leading-order chiral Lagrangian
\cite{GL85}, except that the speed of the mesons in the medium differs from
the speed of light and the Lagrangian contains additional contribution from
the field $\theta$.

    Following the discussions of Refs.~\cite{SS00} and \cite{CG99},
the most general Lagrangian of the diquark kinetic term invariant
under ${\cal G}$ with two derivatives is
\beq
 \label{d-kin-term}
 {\cal L}_d^{\rm kin} = {f_{\tilde \pi}^2} g_{\tilde \pi}^ {\mu \nu }
  {\rm Tr}\left(  \partial _\mu
 \tilde \Delta \partial _\nu \tilde \Delta ^\dag  \right)
  + \frac{f_{\tilde \eta'}^2}{2}{g_{\tilde \eta'}^{\mu \nu } \partial
  _\mu \tilde \theta \partial _\nu \tilde \theta } + \frac{f_{H}^2}{2}
 {g_{H}^{\mu \nu } \partial _\mu \phi \partial _\nu \phi },
\eeq
with the additional metric tensor
$g_{H}^{\mu \nu } = {\rm diag}(1,v_{H}^2,v_{H}^2,v_{H}^2)$.
The generalized pion fields $\tilde \pi^j$ are defined by
\beq
    \tilde \Delta = \exp \left(i\lambda^j {\tilde \pi}^j/f_{\tilde \pi}
    \right).
\eeq
The three terms in Eq.~(\ref{d-kin-term}) are the kinetic terms for pions,
$\eta'$, and the $H$ boson, respectively \cite{SS00,CG99,HLM00}.

    At extremely high density, the decay constants $f_{\tilde \pi}$,
$f_{\tilde \eta'}$, and $f_{ H}$ are found by matching to their microscopic
asymptotic values \cite{SS00}:
\beq
  \label{eq:fpi}
   f_{\tilde {\pi}}^2 \rightarrow \frac {21-8{\rm
  ln}2}{18}\frac{\mu^2}{2\pi^2}, \ \
   f_{H,\tilde {\eta'}}^2 \rightarrow \frac{3}{4}\frac{\mu^2}{2 \pi^2},
\eeq
while the velocities approach the asymptotic sound speed:  $v_{\tilde
{\pi},\tilde {\eta'},H}^2 \rightarrow 1/3$.

\subsubsection{Mass terms}

    We now assume small but finite quark masses and construct the possible
terms in the GL free-energy allowed by the symmetry
group ${\cal G}$, where $M$ is the $3\times3$ quark mass matrix in flavor
space.  In the intermediate density region, where $U(1)_A$ symmetry is
violated by the axial anomaly, the lowest order possible mass terms in the
GL free-energy are ${\cal O} (M)$, unlike at asymptotically high density where
the leading terms are ${\cal O}(M^2)$.

 The bare quark mass term in QCD reads
\beq
{\cal L}_{\rm QCD}^{\rm mass}=\bar q_L M q_R + {\rm h.c.} .
\eeq
As in the standard procedure to build up the chiral Lagrangian, we treat
the matrix $M$ as a spurion field and assume it to transform under ${\cal G}$
as,
\beq
 \label{trans:M}
  M \to e^{-2i\alpha_A}V_L M V_R^\dag.
\eeq
This transformation law together with Eqs.~(\ref{trans:phi}),
(\ref{trans:dLL}) and (\ref{trans:dLR}) enable us to write down the possible
mass terms.  It should be stressed that $\Phi$, $d_L d_R ^\dag$ and $M$ share
the same transformation property under ${\cal G}$ except for the $U(1)_A$
phase rotation.  Therefore, the guiding principle to construct the quark mass
terms is to replace the $\Phi$'s or $d_L d_R ^\dag$'s involved in the GL
potential $\Omega_{{\chi},d,{\chi d}}$, in Eqs.  (\ref{eq:GL-chi}),
(\ref{eq:GL-d}) and (\ref{eq:GL-coup}), with $M$.

    The general mass terms of the chiral and diquark fields obtained from this
procedure are:
\beq
  \label{mass-term1}
  {\cal L}^{\rm mass} =&& A_0 \left[ {\rm Tr} \left( {M\Phi ^\dag  } \right) +
  {\rm h} {\rm .c}. \right]
  \nonumber \\
                    &+& B_0 \left[ \varepsilon _{abc} \varepsilon _{ijk}
  M_{ai} \Phi _{bj} \Phi _{ck}  + {\rm h}{\rm .c}. \right]
  \nonumber \\
                    &+& C_1 \left[ {\rm Tr}\left ( {M\Phi ^\dag  } \right){\rm
   Tr}\left ( {\Phi \Phi ^\dag  } \right)  + {\rm h}{\rm .c}. \right]
  \nonumber \\
                    &+& C_2 \left[ {\rm Tr}\left\{ \left( {M\Phi ^\dag  }
   \right)\left ( {\Phi \Phi ^\dag  } \right) \right\} + {\rm h}{\rm .c}.
  \right]
  \nonumber \\
                    &+& \Gamma_1 \left[ {\rm Tr}
                    \left\{ {M\left( {d_R d_L
    ^\dag  } \right)} \right\}  + {\rm h}{\rm .c}. \right]
   \nonumber \\
                    &+& \Lambda_1 \left[ {\rm Tr}
    \left\{ {\left( {d_L d_L ^\dag   + d_R d_R ^\dag  }
    \right)\left( {M\Phi ^\dag} \right)} \right\}
    + {\rm h}{\rm .c}.\right]
   \nonumber \\
                    &+& \Lambda_2 \left[ {\rm Tr}
    \left( {d_L d_L ^\dag   + d_R d_R ^\dag  }
    \right){\rm Tr}\left( {M\Phi ^
     \dag  } \right) + {\rm h}{\rm .c}. \right] \nonumber \\
                    &+& \Lambda_3 \left[ \varepsilon _ {abc}
                    \varepsilon_{ijk} M_{ai} \Phi _{bj}
    \left( {d_L d_R ^\dag  } \right)_{ck}  + {\rm h}{\rm .c}. \right].
\eeq
Note that the terms proportional to $B_0$ and $\Gamma_1$, which break
$U(1)_A$ symmetry, originate from the axial anomaly, or equivalently, by
instanton-induced interactions.  The leading mass terms, which are linear in
$M$ and have minimal number of the chiral and diquark fields, become
\beq
 \label{mass-term}
  {\cal L}^{\rm mass} = A_0 \left[ {\rm Tr} \left( {M\Phi ^\dag  } \right) +
 {\rm h} {\rm .c}. \right]
 + \Gamma_1 \left[ {\rm Tr}\left\{ {M\left
 ( {d_R d_L ^\dag  } \right) } \right\} + {\rm h} {\rm .c}. \right].
\eeq

\subsubsection{Overall effective Lagrangian}

    We turn now to the overall effective Lagrangian for $\pi$, $\eta'$ and
$H$.  With the parametrizations of the fields defined in
Eqs.~(\ref{parametrization}) and (\ref{new-field}), the GL potential of the
chiral part $\Omega_{\chi}$ in Eq.~(\ref{eq:GL-chi}) and the interaction part
$\Omega_{\chi d}$ in Eq.~(\ref{eq:GL-coup}) reduce to
\beq
 \label{pot-term1}
 {\cal L}_{\chi} &=&   c_0 \sigma ^3 \cos \left( {6\theta } \right)
 + {\rm const.},
  \\
 \label{pot-term3}
  {\cal L}_{\chi d}
  &=&  \gamma _1 d^2 \sigma \left[e^{-2i(2 \tilde \theta + \theta)} {\rm Tr}
  \left(\tilde \Delta^\dag \Sigma \right) + {\rm h.c.} \right]
  + \lambda _3 d^2 \sigma^2 \left[e^{2i (\tilde \theta -\theta)}
  {\rm Tr} \left(\tilde \Delta^\dag \Sigma \right)
  + {\rm h.c.} \right] + {\rm const.},
\eeq
while the diquark part $\Omega_{d}$ in Eq.~(\ref{eq:GL-d}) reduces to a
constant and can be neglected.  From Eqs.~(\ref{chi-kin-term}),
(\ref{d-kin-term}), (\ref{mass-term}), (\ref{pot-term1}) and
(\ref{pot-term3}), the overall effective Lagrangian takes the form up to
${\cal O}(\pi^2, \tilde \pi^2, \eta^{'2}, \tilde {\eta}^{'2})$:
\beq
 \label{eff-Lagrangian2}
  {\cal L}^{\rm eff} &=& {\cal L}^{\rm eff}_ {(\pi)} + {\cal L}^{\rm
  eff}_{(H)} + {\cal L}^ {\rm eff}_{(\eta')}, \\
  \label{pi-Lagrangian} {\cal L}^{\rm eff}_{(\pi)}
  &=& \frac{1}{2} g_\pi ^{\mu \nu } \left( {\partial _\mu  \pi ^j } \right)
  \left( {\partial _\nu  \pi ^j } \right) ^\dag
  + \frac{1}{2}g_{\tilde \pi }^{\mu
  \nu } \left( {\partial _\mu \tilde \pi ^j } \right)
  \left( {\partial _\nu  \tilde \pi ^j } \right)^\dag
 - \frac{1} {2}\left( {\begin{array}{*{20}c}
    \pi  & {\tilde \pi }  \\
  \end{array}}
  \right)M_{\left( \pi \right)} \left( {\begin{array}{*{20}c}   \pi   \\
    {\tilde \pi }  \\
  \end{array}} \right) ,
 \\
  \label{H-Lagrangian}
  {\cal L}^{\rm eff}_{(H)} &=& \frac{1}{2}g_H^
 {\mu \nu } \left( {\partial _\mu H} \right)
  \left( {\partial _\nu  H} \right)^\dag , \\
  \label{eta-Lagrangian}
  {\cal L}^{\rm eff}_{(\eta')} &=&
  \frac{1}{2}g_{\eta '}^{\mu \nu } \left
  ( {\partial _\mu  \eta '} \right)
  \left( {\partial _\nu  \eta '} \right)^\dag
  + \frac{1}{2}g_{\tilde \eta '}^{\mu
  \nu } \left( {\partial _\mu
  \tilde \eta '} \right)
  \left( {\partial _\nu  \tilde \eta'} \right)^\dag
 -\frac{1} {2}\left(
  {\begin{array}{*{20}c}
    {\eta '} & {\tilde \eta '}  \\
  \end{array}} \right)M_{\left( {\eta '} \right)} \left(
  {\begin{array}{*{20}c}
    \eta   \\
    {\tilde \eta '}  \\
\end{array}} \right),
\eeq
where the quark mass matrix is assumed to be diagonal and flavor
symmetric, $M={\rm diag}(m_q,m_q,m_q)$.  The fields are redefined as
\beq
  \theta=\frac{\eta'}{f_{\eta'}}, \ \ \tilde \theta=\frac{\tilde \eta'}
  {f_{\tilde \eta'}}, \ \ \phi=\frac{H}{f_H} .
\eeq
The mass matrices of $(\pi, \tilde{\pi})$ and $(\eta', \tilde{\eta}')$ are
\beq
 \label{matrix-pi}
  M_{\left( \pi  \right)}  &=&
\left( {\begin {array}{*{20}c}
    {\frac{1}{f_\pi  ^2 } (\gamma _1 d^2 \sigma  + \lambda
  _3 d^2 \sigma ^2  + A_0 m_q\sigma)} &
{ -\frac{1}{f_\pi  f_{\tilde \pi } }
(\gamma _1 d^2 \sigma + \lambda _3d^2 \sigma ^2)}
\\
{ -\frac{1}{f_\pi  f_{\tilde \pi }}
(\gamma _1 d^2 \sigma  + \lambda _3 d^2 \sigma ^2)} &
{\frac{1}{f_{\tilde \pi } ^2}
(\gamma _1 d^2 \sigma  + \lambda
  _3 d^2 \sigma ^2  + \Gamma _1 m_q d^2)}
\\
   \end{array}} \right).
  \\
   \label{matrix-eta}
  M_{\left( {\eta '} \right)}  &=&
12\left( {\begin{array}{*{20}c}
    {\frac{1}{f_{\eta '} ^2}(3c_0 \sigma ^3  + 2\gamma _1
  d^2 \sigma  + 2\lambda _3 d^2 \sigma ^2 + 2A_0 m_q \sigma)} &
{\frac{1}{f_{\eta '} f_{\tilde \eta '}}(4\gamma _1 d^2
  \sigma  - 2 \lambda_3 d^2 \sigma ^2)} \\
{\frac{1}{f_{\eta '} f_{\tilde \eta '}}
(4\gamma _1 d^2 \sigma  - 2 \lambda _3 d^2 \sigma ^2)} &
{\frac{1}{f_{\tilde \eta '} ^2}
(8\gamma _1 d^2 \sigma  + 2 \lambda_3 d^2 \sigma ^2
  + 8 \Gamma_1 m_q d^2)}
\\
 \end{array}} \right).
\eeq
Since we have degenerate masses of up, down, and strange quarks,
the Bedaque-Sch\"{a}fer term \cite{BS02} originating from the effective modification
of the chemical potential due to the mass differences does not appear here.
The off-diagonal components of $M_{(\pi)}$
originate from the chiral-diquark coupling.  As the form of $M_{(\pi)}$ shows,
both $\pi$ and $\tilde {\pi}$ are exactly massless in the absence of
$\sigma$-$d$ coupling and quark masses.

\subsection{Mass spectra of low-lying collective modes}
\label{sec:mass_spectra}

    In this subsection, we investigate the mass spectra of pions and $H$ on
the basis of the effective Lagrangian obtained in the last section.  (For
$\eta'$, see Appendix \ref{sec:spectra-eta'}.)  In particular, we obtain a
generalized Gell-Mann-Oakes-Renner (GOR) relation in the intermediate density
region.  We also discuss a possible connection to hadron-quark continuity
for these excited states.

\subsubsection{$H$ boson}

    As we see from Eq.~(\ref{H-Lagrangian}), the $H$ boson is exactly massless
irrespective of the quark masses.  This is because the $H$ boson only exists
in the diquark part and is not affected by the chiral-diquark coupling, so
that there are no mass terms for $H$ involving the quark masses.  On the other
hand, the mass spectra of $\pi$ and $\eta'$, unlike the $H$ boson, are
non-trivial on account of mixing through the chiral-diquark coupling.

\begin{figure}[t]
\begin{center}
\includegraphics[width=8.5cm]{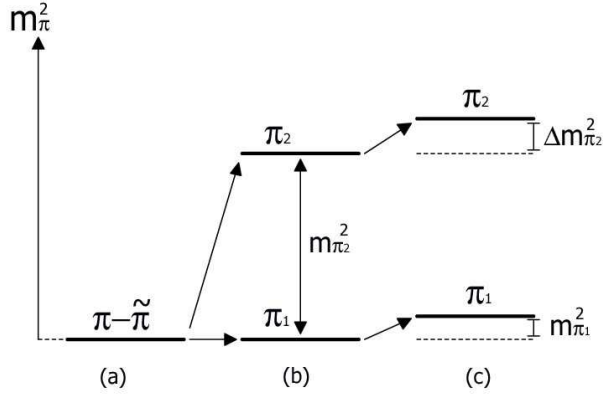}
\end{center}
\vspace{-0.5cm}
\caption{Mass spectra of pions in the intermediate density region:  (a)
without $\sigma$-$d$ mixing ($\gamma= \lambda=0$); (b) with $\sigma$-$d$
mixing ($\gamma \neq 0, \lambda \neq 0$); and (c) in addition with finite
quark masses ($m_q \neq 0$).  Explicit expressions for the $m_{\pi}^2$'s are
given in the text.}
\label{fig:spectra_pi}
\end{figure}

\subsubsection{Pions}

    The eigenstates of the mass matrix $M_{(\pi)} $ in Eq.~(\ref{matrix-pi})
can be written as
\beq
  \label{mixing}
  \left( {\begin{array}{*{20}c}
    {\pi _1 }  \\
    {\pi _2 }  \\
  \end{array}} \right) = \left( {\begin{array}{* {20}c}
    {\cos \vartheta } & {\sin \vartheta }  \\
    { - \sin \vartheta } & {\cos \vartheta }  \\
  \end{array}} \right)\left( {\begin{array}{*{20} c}   \pi   \\
    {\tilde \pi }  \\
  \end{array}} \right),
\eeq
with mixing angle $\vartheta$.  We compute the pion masses $m_{\pi _{1} }$
and $m_{\pi _{2} }$ (we take $m_{\pi _ {1}}^2 < m_{\pi _{2}}^2$) with and
without quark masses. \\
    \noindent (i) In the chiral limit, $m_q=0$, the pion masses are
\beq
  m_{\pi _{1} }^2 (m_q=0)  &=& 0 , \\
 \label{pi2mass}
 m_{\pi _{2} }^2 (m_q=0)
 &=& \left( {\gamma _1 d^2 \sigma  + \lambda _3 d^2 \sigma ^2 } \right)
 \frac{{f_\pi  ^2  + f_{\tilde \pi } ^2 }} {{f_\pi  ^2 f_{\tilde \pi } ^2 }} ,
\eeq
with the mixing angle $\vartheta_0$, which satisfies
\beq
 \label{mixing angle0}
 \tan \vartheta_0  &=& \frac{f_{\tilde \pi}}{f_\pi}.
\eeq
\\

\noindent (ii) For non-vanishing $m_q$,
we find
\beq
\label{mGOR-relation}
 m_{\pi _{1} }^2 &=&
  \frac{m_q}{f_\pi ^2  +  f_{\tilde \pi} ^2 }
    \left( {A_0 \sigma  + \Gamma _1 d^2 }\right),\\
  \label{pi2mass2}
  m_{\pi _{2} }^2 &=& m_{\pi _{2} }^2 (m_q=0) + \frac{m_q}{f_\pi^2
   + f_{\tilde \pi }^2 }\left( \frac{{f_ {\tilde \pi }^2 }}{{f_\pi ^2 }}A_0
  \sigma
  + \frac{{f_\pi ^2 }}{{f_{\tilde \pi }^2 }}\Gamma _1 d^2 \right),
\eeq
with the mixing angle
\beq
 \label{mixing angle}
  \tan \vartheta  = \frac{f_{\tilde
  \pi}}{f_\pi}
   +\frac{f_\pi}{f_{\tilde \pi}}\frac
   {f_\pi^2 \Gamma _1 d^2 - f_{\tilde \pi}^2 A_0 \sigma} {(f_\pi ^2  +
   f_{\tilde \pi }^2 )(\gamma _1 d^2 \sigma  + \lambda _3 d^2 \sigma ^2)}m_q.
\eeq

    The resultant mass spectra of pions is shown in Fig.~\ref{fig:spectra_pi}.
Panel (a) shows the case with neither the chiral-diquark coupling nor finite
quark masses.  In this case both $\pi$ and $ \tilde \pi$ are massless and
degenerate.  Figure~\ref{fig:spectra_pi}(b) shows the effects of the
chiral-diquark coupling without finite quark masses; here the masses are
split, and a massless mode still remains.  Panel (c) shows the effects of both
chiral-diquark coupling and finite quark masses; here because of the explicit
breaking of chiral symmetry by quark masses, the masses of $\pi_1$ and
$\pi_2$ are lifted by $\Delta m_{\pi_1}^2$ and $\Delta m_{\pi_2}^2$
respectively.
    The mass formula for the lightest pseudoscalar meson,
Eq.~(\ref{mGOR-relation}), is a generalized form in dense QCD of the
Gell-Mann-Oakes-Renner (GOR) relation connecting the masses of pseudoscalar
bosons to the chiral and diquark condensates.  The $\Gamma_1$ term, which
originates from the effect of the axial anomaly in the pion Lagrangian, ${\cal
L}^{\rm eff}_{(\pi)}$, shows the crucial role of the axial anomaly not only
for the phase structure but also for the excitation spectra in the
intermediate density region.  When the diquark condensate $d$ decreases as the
density becomes low, Eq.~(\ref{mGOR-relation}) reduces to the standard GOR
relation $f_{\pi}^2 m_{\pi}^2 = A_0 m_q{ \sigma}$.  On the other hand, at
asymptotically high density, the chiral condensate $\sigma$ is small and the
axial anomaly is highly suppressed as $\Gamma_1 \sim \mu(\Lambda_{\rm
QCD}/\mu)^9(1/g)^{14}$ \cite{TS}.  Then, the linear term in $m_q$ disappears
on the right side of Eq.~(\ref{mGOR-relation}) and the leading term becomes
${\cal O}(m_q^2)$, with $m_{\pi_1}^2 \propto m_q^2 d^2$.  This result is
consistent with observations given in \cite{ARW99,SS00} at high density.

    Note that $\pi_2$ is much heavier than $\pi_1$ and it is thus expected to
be radically unstable against decay into $ \pi_1$.  The surviving excitation
mode $\pi_1$ exhibits hadron-quark continuity with increasing baryon density.
The mixing angle $\vartheta=0$ describes the low density limit ($\pi_1=\pi$)
and $\vartheta={\pi}/{2}$ the asymptotically high density limit $(\pi_1=\tilde
\pi)$.  In the intermediate density region, $\pi_1$ is the result of mixing
between the $\sigma \sim \bar q q$ state, as seen in the low density region,
and the $d^2 \sim \bar q \bar q q q$ state, as seen in the high density
region.  The mixing angle $\vartheta$ changes with the chemical potential
$\mu$.  Therefore, across the entire span of the chemical potential, we have
light pseudoscalar modes associated with chiral symmetry breaking, indicating
hadron-quark continuity for the excited pionic states.

\section{summary and conclusion}
\label{sec:summary}

    In this paper, we have investigated the phase structure and excited
spectra of dense QCD in the presence of the axial anomaly.  We constructed the
general GL free-energy and effective Lagrangian for the chiral, diquark
fields, and their interactions, consistent with QCD symmetry, ${\cal G} \equiv
SU(N_f)_L \times SU(N_f)_R \times U(1)_B \times U(1)_A \times SU(3)_C$.

    In Sec.~\ref{sec:phase}, we have shown that the QCD axial anomaly acts as
an external field for the chiral condensate under the influence of the diquark
condensate.  The first order chiral transition is changed to crossover for a
large diquark condensate, and a new critical point driven by the axial anomaly
emerges in the QCD phase diagram.  The former is a realization of hadron-quark
continuity for the QCD ground states.  Determining the precise location of
this new critical point is a future task for phenomenological models and
lattice QCD simulations.  Our schematic phase diagram would be made more
realistic by including effects such as finite quark masses, charge neutrality,
$\beta$-equilibrium, and thermal gluon fluctuations \cite{IMTH,MIHB}.  Open
questions include whether the new critical point would survive in an
inhomogeneous Fulde-Ferrell-Larkin-Ovchinnikov (FFLO) state, and how the COE
phase at low $T$ and low $\mu$, Fig.~\ref{fig:32F}, is affected by quark
confinement.

    In Sec.~\ref{sec:spectra}, we have shown that the massless pions in the
chiral limit in dense QCD result from the mixing of $\pi \sim q \bar q $ and
$\tilde{\pi} \sim qq \bar q \bar q$ states.  Turning on a small but finite
quark mass, we found that the pions acquire a mass both from the chiral
condensate $\sigma$ and the diquark condensate $d^2$, which is summarized in
the generalized Gell-Mann$-$Oakes-Renner formula, Eq.~(\ref{mGOR-relation}).
Here the QCD axial anomaly plays a key role, giving the $d^2$-contribution to
the pion mass.  Our results show explicitly that hadron-quark continuity is
realized not only in the ground state but also in the low-lying collective
excitations.  The continuity of the heavier excitations such as the vector
mesons and baryons remains an interesting problem for future exploration.

    In Sec.~\ref{sec:spectra}, we focused on the COE phase with three
degenerate flavors.  However, in the intermediate density region, in which we
are interested, flavor symmetry breaking due to $m_{\rm s} \gg m_{\rm u,d}$ is
non-negligible and causes mass splittings of the octet of pions in a
non-trivial way.  In particular, it would be interesting to study how, across
the intermediate density region, the inverse mass ordering, $m_{\tilde{\eta}'}
< m_{\tilde{\pi}} \sim m_{\tilde{K}} < m_{\tilde{\eta}}$ \cite{SS00}, at high
density, turns into the normal mass ordering, $m_{\pi} < m_{K} < m_{\eta} <
m_{{\eta}'}$, at low density.

\begin{acknowledgments}

    We would like to thank Mark Alford for helpful discussions and comments.
This research was supported in part by the Grants-in-Aid of the Japanese
Ministry of Education, Culture, Sports, Science, and Technology
(No.~18540253), and in part by NSF Grant PHY03-55014.  Author GB thanks the
University of Tokyo for its kind hospitality as well as support through the
COE.

\end{acknowledgments}

\newpage

\appendix

\section{derivation of phase boundaries and critical points}
\label{sec:phase-diagram}

    In this Appendix we describe more quantitatively the phase boundaries for
three and two massless flavors respectively for both a positive and negative
fourth order term, $b > 0$ or $b < 0$.  We only outline the derivation of the
phase boundaries and critical points; the results can be confirmed explicitly
by comparing the free energies.

\subsection{Three massless flavors}

    To find the minimum of the GL free energy $\Omega_{3F}$ in
Eq.~(\ref{eq:nf3-model}), it is useful to eliminate either of the variables
$\sigma$ or $d$ by solving the stationarity condition,

\beq
\label{eq:d-stationary-condition}
 \frac{\partial\Omega_{3F}}{\partial d} =
  2d \left( \frac{\alpha}{2}+\frac{\beta}{2}d^2-\gamma\sigma \right) =0,
\eeq
which yields $d=0$ and $d^2=(2/\beta)(\gamma \sigma-\alpha/2)$.  The phase
boundaries and the order of the phase transitions are obtained by comparing
the free energies:  $\Omega^{\rm (CSC)}(0,d) (= \alpha d^2/2 + \beta d^4/4$),
$\Omega^{\rm (NG)}(\sigma,0) (= a \sigma^2/2 - c\sigma^3/3 + b\sigma^4/4
+ f\sigma^6/6$), and $\Omega^{\rm (COE)}$ measured with respect to that of the
normal phase, $\Omega^{\rm (NOR)}(0,0)$.  Using the $d \neq 0$ solution of
Eq.~(\ref{eq:d-stationary-condition}), we have
\beq
    \label{eq:effective-sigma-energy}
    \Omega^{\rm (COE)}(\sigma,d(\sigma))
    = -\frac{\alpha^{2}}{4\beta} +\gamma^{*}\sigma+
    \frac{a^{*}}{2} \sigma^2 - \frac{c}{3} \sigma^3 +
   \frac{b}{4}\sigma^4  + \frac{f}{6}\sigma^6 \ \ \ \left(\sigma \ge
   \frac{\alpha}{2\gamma} \right),
\eeq
with
\beq
 \label{def:coefficient_1}
 \gamma^* &\equiv& \alpha \gamma/\beta ,
 \\
 \label{def:coefficient_2}
 a^* &\equiv& a-2\gamma^2/\beta ,
\eeq
or equivalently,
\beq
    \label{eq:effective-d-energy}
    \Omega^{\rm (COE)}(\sigma(d),d)
    = \Omega_0 + \frac{\alpha^{*}}{2}d^2 +
    \frac{\beta^{*}}{4} d^4 + \text{higher order terms} ,
\eeq
with
\beq
 \label{def:coefficient_3}
  \Omega_0 &\equiv& \frac{\alpha^2}{384\gamma^6} \left(48\gamma^4 a -
  16\gamma^3 c \alpha
   +  6\gamma^2 b \alpha^2 + f \alpha^4 \right), \\
 \label{def:coefficient_4}
 \alpha^{*} &\equiv&
  \frac{\beta \alpha}{32\gamma^6 } \left(f\alpha^4 + 4\gamma^2b\alpha^2-
8\gamma^3c\alpha
   + 16\gamma^4a \right), \\
 \label{def:coefficient_5}
  \beta^{*} &\equiv&
  \frac{\beta^2}{32\gamma^6} \left(5f\alpha^4 + 12\gamma^2b\alpha^2-
  16\gamma^3c\alpha +
  16\gamma^4 a- 32\frac{\gamma^6}{\beta}\right).
\eeq

\subsubsection{Positive $b$}
\label{sssec:b-positive-3F}

  For $b>0$, the sixth order $f$
term in Eq.~(\ref{eq:nf3-model}) does not change the qualitative structure of
the phase diagram and can be safely neglected.  In the following, we study the
phase boundaries ABCC', BB', CD, and CDD' in Fig.~\ref{fig:3F} separately and
find the locations of the characteristic points A, B, C and D. \\

\noindent $\bullet $ The critical point A and the first order phase boundary
ABCC'\\

    We first eliminate the $\sigma^3$ term in
Eq.~(\ref{eq:effective-sigma-energy}) for the COE phase by introducing a new
field $\tau = \sigma - c/(3b)$, in terms of which,
\beq
 \label{eq:reduced-effective-sigma-energy2}
 \Omega^{\rm (COE)}(\tau)
 = \Omega_{\text {c}}+\gamma_{\text{c}}^{*}\tau+
   \frac{a_{\text{c}}^{*}}{2} \tau^2 + \frac{b}{4}\tau^4,
\eeq
with
\beq
\label{eq:c-parametrization}
  \Omega_{\text {c}} &=&
  -\frac{\beta}{4\gamma^2}\gamma^{*2}+\frac{c}{3b}\gamma^{*}
  +\frac{c^2}{18b^2}a^{*}-\frac{c^4}{108b^2}  ,
  \nonumber \\
  \gamma_{\text{c}}^{*} &=& \gamma^{*}+\frac{c}{3b}a^{*}-\frac{2c^3}{27b^2} ,
  \\ a_{\text{c}}^{*} &=& a^{*}-\frac{c^2}{3b} ,
  \nonumber
\eeq
where $a^*$ and $\gamma^*$ are given in Eqs.~(\ref{def:coefficient_1}) and
(\ref{def:coefficient_2}).  The system described by
Eq.~(\ref{eq:reduced-effective-sigma-energy2}) is equivalent to an Ising
ferromagnet in an external magnetic field.  Thus the point A corresponds to
the second order critical point of the equivalent magnetic system at
$\gamma_{\text{c}}^{*} = a_{\text{c}}^{*} = 0$.  The location of A in terms
of the original coordinates is:
\beq
\label{point:A}
  \text{A} =
  \left(\frac{c^2}{3b}+\frac{2\gamma^2}{\beta}, - \frac{\beta c^3}{27
   \gamma b^2}\right);
\eeq
thus A is always located in the region $a >0$ and $\alpha <0$, as shown in
Fig.~\ref{fig:3F}.  The first order line AB in the COE phase,
determined by the condition $\gamma_{\text{c}}^{*} = 0$,
\beq
\label{line:AB}
   \text{AB}: \alpha = -\frac{\beta c}{3\gamma b}a + \frac{2\gamma c}{3b}
      +\frac{2\beta c^3}{27\gamma b^2},
\eeq
is straight with negative slope, as shown in Fig.~\ref{fig:3F}.  For
negative $\alpha$, $\tau$ jumps discontinuously across AB between the
solutions in the COE phase, $\tau=\pm \sqrt{-a_{\text {c}}^{*}/b}$.  For
positive $\alpha$, there is a jump between the COE phase and the NOR phase as
shown below.  Thus the coordinates of the point B are
\beq
\label{location3f}
  \text{B} = \left(\frac{2c^2}{9b}+\frac{2\gamma^2}{\beta}, 0\right).
\eeq
The line BC, obtained from $\Omega^{\rm (COE)}=\Omega^{\rm (NOR)}=0$ at
the potential minimum, is first order, since the bracket in
Eq.~(\ref{eq:d-stationary-condition}) always vanishes in the COE phase and
either $\sigma$ or $d$ must be discontinuous from the NOR phase to the COE
phase.  The expression for BC is easy to derive but is too complicated to show
here.  For sufficiently large $\alpha$, the COE phase turns into the NG phase,
as shown below.  The boundary CC' separating the NG phase and the NOR phase is
not affected by the coupling $\gamma$ since $d$ vanishes in both phases.  Then
the condition $\Omega^{\rm (NG)}=\Omega^{\rm (NOR)}=0$ at the potential
minimum implies
 \beq
\label{line:CC'}
    \text{CC'} : a= \frac{2c^2}{9b}.
\eeq

\

\noindent
$\bullet $ The phase boundaries BB' and CD'\\

    To determine the boundaries primarily associated with the diquark
condensate, it is most useful to study the reduced free energy $\Omega^{\rm
(COE)}(\sigma(d),d)$, Eq.~(\ref{eq:effective-d-energy}).  When $\beta^*$ is
positive, the system is equivalent to a usual Ising ferromagnet and shows a
second order transition at $\alpha^*=0$.  One finds two boundaries,
\beq
 \label{eq:3fchiral1}
  \text{BB'} & &: \alpha = 0,\\
  \label{eq:3fchiral2}
  \text{CD'}& & : \alpha = \frac{\gamma}{b}(c+\sqrt{c^2-4ba}).
\eeq
We have chosen the larger root of $b\alpha^2- 2\gamma c \alpha + 4
\gamma^2 a = 0$ obtained from $\alpha^*=0$ in Eq.~(\ref{eq:3fchiral2}) as 
the only root compatible with the condition that $\Omega^{(\rm COE)} =
\Omega^{(\rm NG)} \le \Omega^{(\rm NOR)}$.  Thus the coordinates of the
critical end point C are
\beq
  \text{C} = \left(\frac{2c^2}{9b}, \frac{4c\gamma}{3b}\right) \ \
   {\rm for}\ \beta^* >0 .
\eeq
The condition $\beta^{*}>0$ on the line CD' is equivalent to $0 < \gamma <
(c/3)\sqrt{\beta/b}$.  If $\beta^* < 0$, or equivalently $\gamma >
(c/3)\sqrt{\beta/b}$, a tricritical point D determined by
$\alpha^{*}=\beta^{*}=0$ appears on CD', as shown in the right panel of
Fig.~\ref{fig:3F}.  In this case, the line CD becomes first order and C
becomes a triple point.  The location of D is
\beq
  \text{D} = \left( \frac{c\left( c+c' \right)}{8b}  - \frac{\gamma^2}{\beta},
   \frac{\gamma\left(c+c'\right)}{2b} \right) ,
\eeq
with $c' \equiv \sqrt{c^2+16b\gamma^2/\beta}$.  The point C is determined by 
the coexistence of the three phases, COE, NG, and NOR, so that the free energies
$\Omega^{(\rm COE)}=\Omega^{(\rm NG)}$ measured with respect to the normal
phase, vanish at their minima; then
\beq
  \text{C} = \left( {\frac{{2c^2 }}{{9b}},\sqrt {\frac{\beta }{b}} \left(
  {\frac{c}{{3\sqrt b }} + \frac{\gamma }{{\sqrt \beta  }}} \right)^2 }
   \right)
   \ \  {\rm for}\ \beta^* <0 .
\eeq

\subsubsection{Negative $b$}
\label{sssec:b-negative-3F}

    For $b<0$, the presence of the sixth order $f$-term in
Eq.~(\ref{eq:nf3-model}) is crucial.  Nonetheless, the topological structure
of the phase diagram in Fig.~\ref{fig:3F} is exactly the same as for $b>0$.

\

\noindent
$\bullet $ The critical point A and the first order phase boundary ABCC'\\

    With $b<0$ and the $f$ term, the free energy $\Omega^{\rm
(COE)}(\sigma,d(\sigma))$ in the COE phase has at most three local minima,
which we label $\sigma_i$ ($i=1,2,3$).  The competition among the three phases
leads to at most three first order lines.  However, the detailed stability
analysis outlined below show that there is only one first order line and
critical point in the COE phase.  Thus the phase structure for $b<0$ is the
same as that for $b>0$.

    Let us now determine the two competing chiral condensates in the COE
phase, labelled by $\sigma_1$ and $\sigma_2$, at which $\Omega_1^{\rm (COE)}
=\Omega_2^{\rm (COE)}=\Omega_{\text{min}}$.  Then $\Omega^{\rm
(COE)}(\sigma,d(\sigma)) = -\alpha^2/4\beta +\gamma^{*}\sigma+ a^{*}
\sigma^2/2 - c\sigma^3/3 + b\sigma^4/4 + f\sigma^6/6 =(f/6)(\sigma-\sigma_1)^2
(\sigma-\sigma_2)^2 (\sigma-\tau_1)(\sigma-\tau_2) +\Omega_{\text{min}}$ where
$\tau_{1,2}$ are complex except when $\tau_1=\tau_2=\sigma_3$.  Eliminating
$\tau_{1,2}$ and defining $r \equiv \sigma_1+\sigma_2$ and $s \equiv
\sigma_1\sigma_2$, we have
\beq
\label{relation1}
    & & 4fr^3+(3b-6fs)r-2c = 0,
\eeq
and
\beq
\label{relation2}
   a^{*} &=& fr^4+\frac{b}{2}r^2-fs^2+bs ,
  \nonumber \\
  \gamma^{*} &=& -s r \left( fr^2-fs+\frac{b}{2} \right) ,
  \\
  \Omega_{\rm min} &=& -\frac{f}{6}(2s-3r^2)s^2 - \frac{b}{4}
  -\frac{\alpha^2}{4\beta}. \nonumber
\eeq
In the COE phase, several conditions must be fulfilled:  (i)
$\sigma_{1,2}\ge \alpha/2\gamma$, (ii) $\Omega_1=\Omega_2 \le \Omega_3$, (iii)
$\Omega^{\text {(COE)}} \le \Omega^{\text {(NOR)}}$, and (iv) $\Omega^{\text
{(COE)}} \le \Omega^{\text {(NG)}}$.  From Eq.~(\ref{relation1}), we derive a
necessary and sufficient condition for (i)-(iv), $0 \le 2\sqrt{s} \le r$.
Under this condition, Eq.~(\ref{relation1}), a cubic equation in terms of $r$,
has only one root satisfying $r_{\chi} \le r \le r_{\text{c}}$.  The upper
limit $r_{\chi}$ and lower limit $r_{\text{c}}$, and correspondingly
$s_{\chi}$ and $s_{\text{c}}$, satisfy
\beq
\label{critical point2}
  & & 5fr_{\text{c}}^3+6br_{\text{c}}-4c=0, \ \ s_{\text{c}}= 4r_{\text{c}}^2,
 \\
\label{chiral:bn}
  & & 4fr_{\chi}^3+3br_{\chi}-2c=0,  \ \ s_{\chi}=0.
\eeq
The root $r=r_{\text{c}}$ corresponds to the critical point A since the
two minima merge into one, $\sigma_1=\sigma_2$, while $r=r_{\chi}$ corresponds
to the point B where the two local minima are $\sigma_1=0$ and $\sigma_2 \neq
0$.  Using Eqs.~(\ref{relation1}), (\ref{relation2}) and (\ref{critical
point2}), we thus find
\beq
\label{critical point3}
 {\rm A}  & = & \left(-\frac{3}{8}br_{\text{c}}^2+\frac{3}{4} cr_{\text{c}} +
 \frac{2\gamma^2}{\beta},
 \frac{\beta}{20\gamma} \left( 2br_{\text{c}}^3 -
    3cr_{\text{c}}^2 \right) \right) , \\
 \text{B} & = & \left(-\frac{b}{4}{r_{\chi}}^2 + \frac{c}{2}
 {r_{\chi}} +\frac{2\gamma^2}{\beta}, 0 \right).
\eeq
It can be shown that the point A is always in the fourth quadrant ($a>0$
and $\alpha<0$) and that the point B is always on the $a$-axis.

    The line AB located in $\alpha \le 0$ corresponds to the interval,
$r_{\chi} \le r \le r_{\text{c}}$.  Thus there is only one first order line,
AB, and critical point, A, in the COE phase.  The line BC, whose explicit form
is complicated, is first order, as for $b>0$, since
Eq.~(\ref{eq:d-stationary-condition}) is always satisfied in the COE phase and
either $\sigma$ or $d$ is discontinuous across BC.  The boundary CC' is not
affected by the diquark condensate and, following the similar argument leading to
Eq.~(\ref{relation1}), we have
\beq
 \text{CC'}: a= -\frac{b}{4}{r_{\chi}}^2 + \frac{c}{2}{r_{\chi}} ,
\eeq
which is in the region $a>0$.

\

\noindent
$\bullet $ Phase boundaries BB' and CD'\\

    As for $b>0$, the lines BB' and CD' obtained from $\alpha^{*}=0$ are
\beq
 \label{eq:3fchiral3}
 \text{BB'}:& & \alpha = 0 ,\\
 \label{eq:3fchiral4}
 \text{CD'}:& & f\alpha^4 + 4\gamma^2 b\alpha^2 - 8\gamma^3 c \alpha + 16
 \gamma^4 a = 0.
\eeq

    The maximum root of $\alpha$ in Eq.~(\ref{eq:3fchiral4}) should be chosen
to satisfy the condition on CD':  $\Omega^{(\rm COE)} = \Omega^{(\rm NG)} \le
\Omega^{(\rm NOR)}$.  The coordinates of C for $\beta^{*}> 0$ are easily found
from the crossing of CD' and CC':
\beq
  \text{C} = \left(-\frac{b}{4}{r_{\chi}}^2 + \frac{c}{2}
   {r_{\chi}} , 2\gamma r_{\chi} \right), \ \
     {\rm for}\ \beta^* >0 .
\eeq
If $f<b^3({c-3c''})/{(c+c'')^3}$ with $c'' \equiv
\sqrt{c^2-8{b\gamma^2}/{\beta}}$, $\beta^{*}$ can change sign on the line CD',
and a tricritical point D determined by $\alpha^{*}=\beta^{*}=0$ appears, as
shown in the right panel of Fig.~\ref{fig:3F}.  In this case, the line CD
becomes first order and the C becomes a triple point.

    Thus we find that the topological phase structure in the massless
three-flavor system with $\gamma>0$ and $\lambda=0$ is independent of the sign
of $b$, and a critical point A always appears in the COE phase as shown in
Fig.~\ref{fig:3F}.

\subsubsection{Effect of small $\lambda$}

    Finally, we discuss the effect of the term $\lambda d^2 \sigma^2$,
hitherto neglected.  With this term, the stationarity condition becomes [cf.
Eq.~(\ref{eq:d-stationary-condition})],
\beq
  \label{eq:d-stationary-condition2}
  \frac{\partial\Omega_{3F}}{\partial d^2}
  =\frac{\alpha}{2}+\frac{\beta}{2}d^2-\gamma\sigma+\lambda\sigma^2=0.
\eeq
Accordingly, the reduced free-energy $\Omega^{\rm (COE)}
(\sigma,d(\sigma))$ in Eq.~(\ref{eq:effective-sigma-energy}) is modified as
\beq
   \label{eq:effective-sigma-energy2}
   \Omega^{\rm (COE)}(\sigma,d(\sigma))
   = -\frac{\alpha^{2}}{4\beta} +\gamma^{*}\sigma+
   \frac{a^{*}_{\lambda}}{2} \sigma^2 - \frac{c_{\lambda}}{3} \sigma^3
   + \frac{b_{\lambda}}{4}\sigma^4  + \frac{f}{6}\sigma^6 ,
\eeq
where the new coefficients are
\beq
a^*_{\lambda} = a^*- 2\alpha \frac{\lambda}{\beta}, \ \ \
c_{\lambda} = c - 6\gamma \frac{\lambda}{\beta}, \ \ \
b_{\lambda} = b- 4 \lambda \frac{\lambda}{\beta}.
\eeq
Therefore, insofar as $|\lambda / \beta| \ll 1$, as is suggested
microscopically (see Appendix \ref{sec:sign-lambda}),
the $\lambda$ term does not qualitatively
change the phase diagram in Fig.~\ref{fig:3F}.

    \subsection{Two massless flavors} Following the same argument as in the
three flavor case, we use the stationarity condition for the GL free energy
$\Omega_{2F}$ in Eq.~(\ref{eq:nf2-model}),
\beq
\label{eq:d-stationary-condition2}
 \frac{\partial\Omega_{2F}}{\partial d} =2d
 \left( \frac{\alpha}{2}+\frac{\beta}{2}d^2+\lambda \sigma^2 \right) =0,
\eeq
to obtain the reduced free-energies:
\beq
  \label{eq:effective-sigma-energy2}
  \Omega^{\rm (COE)}(\sigma,d(\sigma))
  = -\frac{\alpha^{2}}{4\beta} +
   \frac{a'}{2} \sigma^2 + \frac{b'}{4}\sigma^4 + \frac{f}{6}\sigma^6,
\eeq
for $\sigma^2 \leq - \alpha/(2\lambda)$, and
\beq
  \label{eq:effective-d-energy2}
  \Omega^{\rm (COE)}(\sigma(d),d)
  = \Omega_0 + \frac{\alpha '}{2}d^2 +
  \frac{\beta '}{4} d^4 - \frac{f \beta^3}{48\lambda^3}d^6,
\eeq
for $d^2 \leq - \alpha/\beta$.  Here the coeficients are
\beq
  a'&\equiv& a-2\lambda \alpha/\beta ,
   \nonumber \\
   b'&\equiv& b-4\lambda^2/\beta ,
  \nonumber \\
   \Omega_{2F}^0 &\equiv& \alpha \left(-12\lambda^2 a +3\lambda b \alpha -f
   \alpha^2 \right)/(48\lambda^3) ,
   \\
   \alpha' &\equiv& -(f\alpha^2-2\lambda b
   \alpha+4\lambda^2a)\beta^2/(8\lambda^3) ,
  \nonumber \\
  \beta' &\equiv& \left(b' \lambda-f\alpha \right)\beta^2/(4\lambda^2) .
  \nonumber
\eeq
Note that the potential (\ref{eq:effective-d-energy2}) is bounded, despite
the negative $d^6$ term, a consequence of the condition $d^2 \leq -
\alpha/\beta$.

\subsubsection{positive $b$}

    As in an Ising ferromagnet, the phase boundaries, between NOR and NG, and
NOR and CSC, for $b>0$ and $f=0$ are characterized by second order lines,
$a=0$ and $\alpha=0$, respectively.  For both $\beta'>0$ and $b'>0$, or
equivalently,
\beq
 \frac{1}{2}\sqrt{b\beta}>\lambda>0,
\eeq
the COE-NG and COE-CSC phase boundaries are characterized by $\alpha'=0$
and $a'=0$, respectively.  These conditions can be rewritten as
\beq
 \alpha & =&  \frac{2a\lambda}{b},  \ \ \ {\rm \ COE-NG \ boundary}  , \\
 \alpha  & =&  \frac{a\beta}{2\lambda},\ \ \ {\rm \ COE-CSC \ boundary} ,
\eeq
depicted in the left panel of Fig.~\ref{fig:2Fp}.

    On the other hand, when $\lambda > \sqrt{b\beta}/2$,
the COE phase ceases to exist. Then the
competition between the free energies $\Omega^{\rm (NG)}_{\rm min} =
-a^2/(4b)$ and $\Omega^{\rm(CSC)}_{\rm min}= -\alpha^2/(4\beta)$ determines
the first order NG-CSC boundary:
\beq
  \alpha=a\sqrt{\beta/b},
\eeq
as shown in the right panel of Fig.~\ref{fig:2Fp}.

\subsubsection{negative $b$}

    The case $b<0$ requires introduction of a positive $f$ term.\\

\noindent
$\bullet $ First-order phase boundary  E'EFF'\\

    The phase boundaries EE' and FF' are determined by $\Omega^{\rm (NG)} =
\Omega^{\rm (NOR)}$ and $\Omega^{\rm (COE)} = \Omega^{\rm (CSC)}$ at the
potential minimum,
\beq
  \text{EE'} : & & a = 3b^2/(16f).
  \\
  \text{FF'} : & &
  \alpha=\frac{\beta}{2\lambda}\left[a-3\left( b-4\lambda^2/\beta
  \right)^2/(16f) \right].
\eeq
To determine the boundary EF, we use the free-energies of the NG and CSC
phases,
\beq
 \Omega^{\rm (NG)}_{\rm min} &=&
 -{\left[6bfa+(b^2-4fa)^{3/2}-b^3\right]} \mathord{\left/
 {\vphantom {{\left[6bfa+(b^2-4fa)^{3/2}-b^3\right]}
   {(24f^2)}}} \right. \kern-\nulldelimiterspace}{(24f^2)},
\nonumber \\
\Omega^{\rm(CSC)}_{\rm min} &=& -\alpha^2/(4\beta),
\eeq
where the global minima at $\sigma \neq 0$ in the NG phase and $d \neq 0$
in the CSC phase implicitly ensure $b^2-4fa \ge 0$ ($\alpha \le 0$).  From
$\Omega^{\rm (NG)}_{\rm min} = \Omega^{\rm(CSC)}_{\rm min}$, we obtain
\beq
 \text{EF}: \alpha=-\left \{ \beta{\left[6bfa+(b^2-4fa)^{3/2}-b^3\right]}
 \mathord{\left/
 {\vphantom {{\left[6bfa+(b^2-4fa)^{3/2}-b^3\right]} {(6f^2)}}} \right.
   \kern-\nulldelimiterspace}{(6f^2)} \right \}^{1/2}.
\eeq
The critical end points E and F are then
\beq
 \label{coordinate:EF}
 \text{E} = \left(\frac{3b^2}{16f}, 0\right),
 \ \ \text{F} = \left(\frac{3b'}{16f} \left({b'} +
 \frac{16\lambda^2}{\beta}\right), \frac{3\lambda b'}{2f}\right).
\eeq
\

\noindent
$\bullet $ Second order phase boundary EG' and FH'\\

    The boundary EG' is not affected by the chiral condensate and is
characterized by $\alpha=0$, while FH' is given by $\alpha'=0$:
\beq
 \text{EG'}& & : \alpha=0, \\
 \text{FH'}& & : \alpha=\frac{\lambda}{f}\left( b-\sqrt{b^2-4fa} \right).
\eeq
These boundaries indeed terminate at the points E and F given in
Eq.~(\ref{coordinate:EF}).  On EG' and FH', the coefficient $\beta'$ never
changes sign, and therefore a tricritical point does not appear, unlike in the
three-flavor case (the point D in the right panel of Fig.~\ref{fig:3F}).

    While we have left out details of the derivation of the phase structure in
Fig.~\ref{fig:2Fn}, the structure can, as mentioned earlier, be explicitly
checked by direct comparison of the free energies.

\section{sign of $\gamma$ for three flavors}
\label{sec:sign-gamma}

    In this Appendix we summarize how the anomaly-induced interaction $-\gamma
d^2\sigma$ in Eq.~(\ref{eq:nf3-model}) results from the basic 6-Fermi
interaction.  The latter has the form \cite{HK94}
\beq
  {\cal H}_{\rm inst}
   = - {\cal L}_{\rm inst}
   = - g_{\rm D} \left[ \det_{i,j} \bar{q}_a^i (1-\gamma_5) q_a^j +{\rm h.c.}
  \right]
   = - 2g_{\rm D}  \left[ \det_{i,j} \bar{q}_{Ra}^i q_{La}^j +{\rm h.c.}
  \right],
\eeq
with $g_{\rm D}<0$.  The final ($c_0$) term of the GL free energy in
Eq.~(\ref{eq:GL-chi}) originates from the expectation value (in lowest order)
of ${\cal H}_{\rm inst}$; inserting $\Phi$ from Eq.(\ref{eq:phi-def}) into
Eq.(\ref{eq:GL-chi}), we have
\beq
 \label{eq:6-fermi}
 {\Omega}_{\rm inst} =\frac{c_0}{2}   \ G_{\chi}^{3}\
 \big\langle  \det_{i,j}  (\bar{q}_{Ra}^i q_{La}^j)+{\rm h.c.}\big\rangle
  =  \frac{c_0}{2} \cdot G_{\chi}^{3}
  \cdot \frac{1}{6} \cdot \epsilon_{ijk} \epsilon_{i'j'k'}
  \langle (\bar{q}_{Ra}^i q_{La}^{i'})(\bar{q}_{Rb}^j
  q_{Lb}^{j'})(\bar{q}_{Rc}^k q_{Lc}^{k'}) \rangle
  + {\rm c.c.}.
\eeq
Thus the GL coefficient is related to the microscopic coupling by $c_0
G_{\chi}^3/2 = - 2 g_{\rm D} > 0$.  The coefficient $c_0$ must be positive in
order for $m_{\eta'}^2 \sim c_0 $ to be positive (see Eq.~(\ref{eta'-mass0})).
In addition, for $c_0>0$, the chiral condensate at low temperature is
positive.  We have $G_{\chi} > 0$ as well.

    To proceed, we replace one of the $\bar{q}_R q_L$ pairs in
Eq.~(\ref{eq:6-fermi}) by the chiral field $\Phi$, which gives
\beq
\label{eq:gam2}
   {\Omega}_{\rm inst}  &\simeq& - \frac{c_0}{2} \cdot G_{\chi}^{2} \cdot
  \frac{3}{6}\cdot \epsilon_{ijk} \epsilon_{i'j'k'}
  \langle (\bar{q}_{Ra}^i q_{La}^{i'})(\bar{q}_{Rb}^j q_{Lb}^{j'}) \rangle
  \Phi_{k'k} + {\rm c.c.}  \\
  \label{eq:gam3}
  &\simeq&  \frac{c_0}{2} \cdot G_{\chi}^{2}
  \cdot \frac{3}{6} \cdot \frac{2}{4} \cdot \epsilon_{ijk} \epsilon_{i'j'k'}
   \langle (\bar{q}_{Ra}^i C \bar{q}_{Rb}^j)(q_{Lb}^{j'} C
   q_{La}^{i'})\rangle  \Phi_{k'k} + {\rm c.c.}   \\
   \label{eq:gam4}
  &=& c_0 \left( \frac{G_{\chi}}{G_d} \right)^2
  {\rm Tr} [(d_{_R} d_{_L}^{\dagger}) \Phi]  + {\rm c.c.}
  \equiv \gamma_1  {\rm Tr} [(d_{_R} d_{_L}^{\dagger}) \Phi]  + {\rm c.c.}  .
\eeq

    In deriving (\ref{eq:gam3}) from (\ref{eq:gam2}) we use the fermion
anti-commutator twice and the Fierz transform; The overall sign changes from
$-$ to $+$ because the coefficient of the $CC$ term in the Fierz transform is
negative.  Then in deriving (\ref{eq:gam4}) from (\ref{eq:gam3}) we use the
definition of the diquark condensate, Eq.~(\ref{phi-dl}).  The contraction of
the $\epsilon$-tensor three-times gives $+(-)^2$ preserving the overall sign.
The coefficient $\gamma_1$, defined in Eq.~(\ref{eq:GL-coup}) is thus $c_0
(G_{\chi}/G_d)^2 >0$.

    With the CFL ansatz (\ref{CFL ansatz}), the anomaly-induced term reduces
to
\beq
  {\Omega}_{\rm inst} \rightarrow - 3 \gamma_1 d^2 \sigma \equiv - \gamma d^2
 \sigma.
\eeq
The minus sign originates from $d_L = - d_R$.  Thus we finally find that
$\gamma$ [Eq.~(\ref{eq:nf3-model})] is given by
\beq
   \gamma = 3 \gamma_1 = 3 c \left( \frac{G_{\chi}}{G_d} \right)^2 > 0.
\eeq
This result is consistent with a similar result given in \cite{TS}, where
it is found that the effect of the quark mass combined with the axial anomaly
reduces the energy of the CFL phase.

\section{$\lambda$, $\beta$ and $b$ from the NJL model}
\label{sec:sign-lambda}

    The $\lambda$ in Eq.~(\ref{eq:nf3-model}) may be evaluated in the
three-flavor and two-flavor NJL models as follows.  We start with the free
energy Eq.~(5.15) in Ref.~\cite{NJL-model} for three flavors, and Eq.~(4.50)
for two flavors, in the chiral limit:
\beq
 \Omega_{3F}(\sigma,d)
 &=& \frac{3\sigma^2}{2G_{\chi}} - c_0 \sigma^3
 + \frac{ 3d^2 }{ 2G_{d} }
 - \int \frac{ d^3 p }{ (2\pi)^3 }  \sum\limits_ \pm
 \left[ 8\omega_8^{\pm} + \omega_1^{\pm} + 16T \ln \left( 1 + e^{ -
 \omega_8^{\pm} /T} \right)
+2T \ln \left( 1 + e^{ - \omega_1^{\pm} /T } \right) \right],
 \label{eqn:3omega} \\
\Omega_{2F}(\sigma,d)
&=& \frac{ \sigma^2 }{ 2G_{\chi} }
+ \frac{ d^2 }{ 2G_{d} }
- \int \frac{ d^3 p }{ (2\pi)^3 } \sum\limits_ \pm
\left[ 4 \omega^{\pm} + 2 E^{\pm}
+ 8T \ln \left( 1 + e^{ -\omega^{\pm} /T} \right)
+ 4 T \ln \left( 1 + e^{ - E^{\pm} /T}      \right) \right] ,
\label{eqn:2omega}
\eeq
where
\beq
E^{\pm} &=& \sqrt{p^2 + M^2} \pm \mu \ \ \ {\rm with } \ \ \
M = \left\{ {\begin{array}{*{20}c}
    \sigma-c_0G_{\chi}\sigma^2/2 \ \ \ ({\rm three\,\, flavors}) \\
   \sigma \ \ \ \ \ \ \ \ \ \ \ \ \ \ \ ({\rm two\,\, flavors})  \\
\end{array}} \right. \\
  \omega_{8(1)}^{\pm} &=& \sqrt{(E^{\pm})^2 + d_{8(1)}^2}, \\
  \omega^{\pm} &=& \sqrt{(E^{\pm})^2 + d^2}.
\eeq

    Here $d_8=d$ is the octet gap, and $d_1 =2d_8$ the singlet gap, for nine
quasi-quarks (three colors times three flavors).  The $c_0$-term originates
from the instanton-induced interaction; we ignore in Eq.~(\ref{eqn:3omega})
the contribution from the quark$-$quark and quark$-$anti-quark six-point
instanton-induced interaction (the $\gamma_1$-term in Eq.~(\ref{eq:GL-coup})).
Since the fermion integral does not diverge in the infrared,
a naive power-series expansion of $\Omega$ in $d$ and $\sigma$ should be valid.

    The gap is determined by minimizing $\Omega$ with respect to $d^2$:
\beq
  \frac{\partial \Omega_{3F}}{\partial d^2}
  &=& \frac{3}{2G_d}- \sum\limits_\pm \int \frac{ d^3 p }{ (2\pi)^3 }
  \left\{ \frac{4}{ \omega_8^{\pm} } \tanh \frac{ \omega_8^{\pm} }{2T}
  + \frac{2}{ \omega_1^{\pm} } \tanh \frac{ \omega_1^{\pm} }{2T} \right\} ,
  \label{eqn:3GED}
 \\
 \frac{\partial \Omega_{2F}}{\partial d^2}
 &=& \frac{1}{2 G_d}-  \sum\limits_\pm \int \frac{ d^3 p }{ (2\pi)^3 }
 \left\{ \frac{2}{ \omega^\pm } \tanh \frac{ \omega^\pm }{2T} \right\} ,
\label{eqn:2GED}
\eeq
where we introduce an ultraviolet cutoff at $p=\Lambda$ to regulate the
integrals.  In the NJL model, $\Lambda \simeq 1 $GeV, while in weak-coupling
QCD at high density, the momentum dependence of the gap function leads to
$\Lambda \simeq 256 \pi^4 (2/3g^2)^{5/2} \mu \gg \mu$ \cite{IMTH}.  In the
following, we always assume $\Lambda \gg \mu$ and $\Lambda \gg T$.  The $-$
terms are the contribution from the particles and the $+$ terms from
anti-particles.  The critical temperature $T_c$ of the super-to-normal
transition for three flavors is determined by setting the right side of
Eq.~(\ref{eqn:3GED}), and of Eq.~(\ref{eqn:2GED}) for two flavors, to zero.

    The term $\lambda \sigma^2 d^2$ in the GL expansion of $\Omega(\sigma, d)$
is obtained as
\beq
 \lambda_{3F} = \left.  \frac{\partial^2 \Omega_{3F}}{\partial
 {\sigma}^2 \partial d^2 } \right|_{\sigma=d=0}
 = - \frac{3}{2\pi^2} \sum_{\pm}  \int_0^{\Lambda}
  dp \ p \frac{\partial}{\partial p} \left[ \frac{1}{p \pm \mu} \tanh
  \frac{ (p\pm \mu)}{2T} \right] .
\label{eq:lambda-int-3}
\eeq
Evaluating the above integral by partial integration, making the
approximation, $\tanh x \simeq 1 \ (x>1)$,
$\tanh x \simeq -1 \ (x < - 1)$ and
$\tanh x \simeq x \ (-1 \le x \le 1)$, we arrive at
\beq
\lambda_{3F} \simeq
 \frac{3}{2\pi^2}  \ln \left( \frac{\Lambda}{2T} \right)^2  > 0 ,
\ \ \lambda_{2F} = \lambda_{3F}/3 .
\label{eq:lambda-integral}
\eeq

    We can similarly calculate the coefficient $\beta$ in the term
$\beta(d^2)^2/4$.  For three flavors
\beq
    \beta_{3F} = 2 \left.  \frac{\partial^2 \Omega_{3F}}{\partial d^2 \partial
 d^2 } \right|_{\sigma=d=0} = -\frac{6}{\pi^2} \sum_{\pm} \int_0^{\Lambda}
 dp \ \frac{p^2}{p\pm \mu } \ \frac{\partial}{\partial p} \left[\frac{1}{p \pm
 \mu} \tanh \frac{ (p\pm \mu)}{2T} \right] ,
\label{eq:beta-int3}
\eeq
which leads to
\beq
\beta_{3F} \simeq \frac{6}{\pi^2}
\left[ \left( \frac{\mu}{2T} \right)^2
+ \ln \left( \frac{\Lambda}{2T} \right)^2
\right]  > 0,
\ \ \beta_{2F}= \beta_{3F}/6.
\label{eq:beta-integral}
\eeq

    For typical values in the NJL model at intermediate density:  $\Lambda =
1$ GeV, $\mu =500$ MeV and $T = 50$ MeV, we have $\lambda_{3F}/\beta_{3F}
\simeq 0.7/18 = 0.04$.  Also, at asymptotically high density in QCD with
$\Lambda \propto \mu$, we have $\lambda_{3F}/\beta_{3F} \sim \ln(\mu/T)^2
/(\mu/T)^2$ which is parametrically small at large $\mu$.

    The coefficient $b$ in the term $b(\sigma^2)^2/4$ can be calculated from
the free energy Eq.~(\ref{eqn:3omega}), or Eq.~(\ref{eqn:2omega}), with $d=0$:
 \beq
  b_{_{N_f}} &=&  2 \left. \frac{\partial^2 \Omega_{_{N_f}}}
  {\partial \sigma^2 \partial \sigma^2 }
 \right|_{\sigma=0} =
 - \frac{3N_f}{4\pi^2} \sum_{\pm}  \int_0^{\Lambda}  dp
  \ p \frac{\partial}{\partial p}
  \left[ \frac{1}{p} \tanh \frac{(p\pm \mu)}{2T} \right]  ,
\label{eq:b-int}
\eeq
where we have assumed $M=\sigma$ for both two and three flavors, for
simplicity.  Evaluating the integral, we find
\beq
 b_{_{N_f}}
 &\simeq&  \frac{3N_f}{4\pi^2}
   \left[
      \left( 1 + \frac{\mu}{2T} \right)
              \ln \left| \frac{\Lambda/2T}{1+\mu/2T} \right|
     +\left( 1 - \frac{\mu}{2T} \right)
              \ln \left| \frac{\Lambda/2T}{1-\mu/2T} \right|
   \right] .
\label{eq:b-integral}
\eeq
Unlike for the coefficients $\lambda$ and $\beta$, the sign of $b$ changes
from positive to negative as $\mu$ increases from zero.

    Finally, we show the integral form of the coefficient $\tilde{b}_{_{2F}}$
which appears in the term, $\tilde{b}_{_{2F}} d^2\sigma^4 /4$ in the GL
expansion of the two flavor NJL model:
 \beq
  \tilde{b}_{_{2F}} &=&  2 \left. \frac{\partial^3 \Omega_{_{2F}}}
  {\partial \sigma^2 \partial \sigma^2 \partial d^2 } \right|_{\sigma=d=0} =
  - \frac{1}{2\pi^2} \sum_{\pm}  \int_0^{\Lambda}  dp
  \ p^2 \left( \frac{1}{p} \frac{\partial}{\partial p} \right)^2
  \left[ \frac{1}{p \pm \mu } \tanh \frac{(p\pm \mu)}{2T} \right]  ,
\label{eq:bp-integral}
\eeq
where the sign of $\tilde{b}_{_{2F}}$ changes, as one can show, from negative to
positive as $\mu$ increases from zero.

\section{Mass spectra of the $\eta'$ meson}
\label{sec:spectra-eta'}

\begin{figure}[h]
\begin{center}
\includegraphics[width=8.5cm]{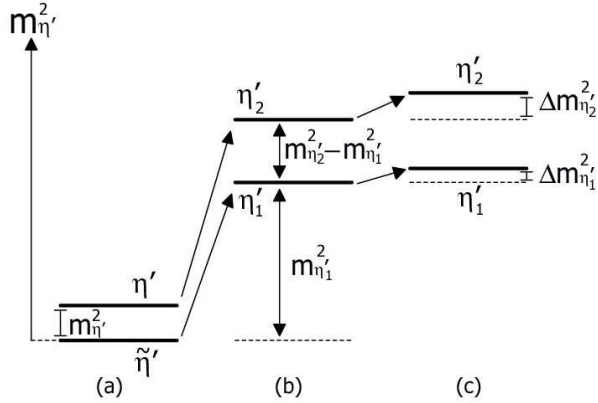}
\end{center}
\vspace{-0.5cm}
\caption{Mass spectra of $\eta'$ meson in the intermediate density region
when (a) $\sigma$-$d$ mixing does not exist ($\gamma= \lambda=0$); (b)
$\sigma$-$d$ mixing is considered ($\gamma \neq 0, \lambda \neq 0$); (c) a
finite quark mass is considered ($m_q \neq 0$) in addition.  Here
$m_{\eta'}^2=36c_0 \sigma^3/{f_{\eta '} ^2 }$, $m_{\eta '_{1,2} }^2(m_q=0) =
12\left( {x_1 \mp {\sqrt {x_1^2 - x_2 } } } \right) $ and $\Delta m_{\eta
'_{1,2} }^2 = 12\left( {x_3 \mp ({{x_1 x_3 - x_4 }})/{{\sqrt {x_1^2 - x_2 }
}}} \right)m_q $. }
\label{fig:spectra_eta}
\end{figure}

    The $\eta'$ mass spectra can be found, similarly to that of the pions, by
diagonalizing the mass matrix Eq.(\ref{matrix-eta}).
Figure~\ref{fig:spectra_eta} shows the mass spectra for $\eta'$:  (a) with
neither the $\sigma$-$d$ coupling nor a quark mass; (b) with the $\sigma$-$d$
coupling but without the quark mass; and (c) with both the $\sigma$-$d$
coupling and non-zero $m_q$.  In case (a), $\tilde \eta'$ is massless, since
there is no term that breaks the $U(1)_A$ symmetry in Eq.~(\ref{eq:GL-d}); on
the other hand, $ \eta'$ is massive because of the axial anomaly, with
\beq
 \label{eta'-mass0}
  m_{\eta'}^2=36c_0 \frac{\sigma^3} {f_{\eta '} ^2 }.
\eeq
This result differs notably from that for pions, where both $\pi$ and
$\tilde{\pi}$ are massless in the absence of the $\sigma$-$d$ coupling and the
quark mass.

    With the $\sigma$-$d$ coupling alone (b), the coupling mixes $\eta'$ with
$\tilde{\eta}'$ and makes the mass eigenstates $\eta'_{1,2}$ much heavier than
in the absence of the coupling.  We have
\beq
   \label{eta'-mass}
   m_{\eta '_{1,2} }^2(m_q=0)
   & = & 12\left( {x_1  \mp {\sqrt
   {x_1^2  - x_2 } } } \right) ,
\eeq
where
\beq \label{def:x12}
  x_1 & = & m_{\eta'}^2 +
    \frac{{ \gamma _1
   d^2 \sigma  + \lambda _3 d^2 \sigma ^2 }}{{f_ {\eta '}^2 }} +
   \frac{{4\gamma_1 d^2 \sigma  + \lambda _3 d^2 \sigma ^2 }}{{f_{\tilde
  \eta'}^2 }},
 \nonumber \\
    x_2 & = & 6\frac{{4c_0 \gamma _1 d^2 \sigma ^4  + c_0 \lambda _3 d^2
  \sigma^5
     + 6\gamma _1 \lambda _3 d^4 \sigma ^3 }}{{f_ {\eta '}^2 f_{\tilde
   \eta'}^2 }}.
\eeq
The $\gamma_1$-term not only mixes the mass spectra of $\eta'$ and $\tilde
\eta'$, but also serves itself as a mass term for $\eta'$ and $\tilde \eta'$.
On the other hand, the $\lambda_3$ term, which preserves $U(1)_A$ symmetry,
acts only to mix $\eta'$ and $\tilde \eta'$.

    In case (c), the non-zero quark mass increases both the $\eta'_1$ and
$\eta'_2$ masses by
\beq
 \label{eta'-mass-correction}
   m_{\eta '_{1,2} }^2  &=&
    m_{\eta '_{1,2} }^2(m_q=0)
  +  12\left
  ({x_3  \mp \frac{{x_1 x_3  - x_4 }}{{\sqrt {x_1^2
   - x_2 } }}} \right)m_q ,
\eeq
with
\beq
 \label{def:x34}
   x_3 & = & \frac{{A_0 \sigma }}{{f_{\eta '}^2 }}
   + \frac{{4\Gamma _1 d^2 }}{{f_{\tilde \eta '}^2 }} ,
  \nonumber \\
  x_4 & = & 2\frac{{6c_0 \Gamma _1 d^2 \sigma ^3  +
  4\Gamma _1 d^2 \left( {\gamma _1 d^2 \sigma
  + \lambda _3 d^2 \sigma ^2 } \right) +
  A_0 \sigma \left( {4\gamma _1 d^2 \sigma
  + \lambda _3 d^2 \sigma ^2 } \right)}}{{f_
  {\eta '}^2 f_{\tilde \eta '}^2 }}.
\eeq
For small $m_q$, $\tilde{\eta}_1'$ becomes heavy owing to the $\sigma$-$d$
coupling.  By contrast, the pion $\pi_1$ remains light, even in the presence
of the coupling.

\end{document}